\documentclass{ROMSOC}
\usepackage[printonlyused,nohyperlinks]{acronym}
\usepackage[draft,textwidth=1.7cm, textsize=footnotesize,color=ROMSOCblue!30,colorinlistoftodos]{todonotes}   % notes showed
\usepackage[sort=use]{glossaries}
\makeglossaries

\newglossaryentry{PRIIP}
{
    name=Packaged retail investment and insurance-based products,
    description={Packaged retail investment and insurance-based products are a broad class of financial instruments that are provided to customers in the EU through banks or financial institutions, which include stocks, bonds, insurance policies, structured funds, structured deposits, and structured products}
}

\newglossaryentry{KID}
{
    name=key information document,
    description={A key information document is a 3-page document provided to the customers by financial institutions for the invested product. It includes the risk and reward profile of the product, the cost of the product, the recommended holding period, possible outcomes, etc., so that investors can understand the product easily.}
}

\newglossaryentry{bank}
{
    name=bank account,
    description={A bank account is an interest-bearing account at a bank to earn interest on the invested amount}
}

\newglossaryentry{shortrate}
{
    name=short-rate,
    description={Short rate is the rate at which bank account grows}
}

\newglossaryentry{CashFlow}
{
    name=cash flow,
    description={Cash flow is the net amount of cash being transferred into or out of a bank account}
}

\newglossaryentry{PV}
{
    name=present value,
    description={The present value is the current value of a stream of cash flows for a given bank account with a specified rate of returns}
}

\newglossaryentry{DF}
{
    name=discount factor,
    description={
A factor which, when multiplied by the predicted future value of the bank account gives its present value}
}

\newglossaryentry{YC}
{
    name=yield curve,
    description={The yield curve is the curve showing interest rates plotted against different maturities for the same financial instrument}
}

\newglossaryentry{maturity}
{
    name=maturity,
    description={The maturity date is the final payment date of a financial instrument}
}

\newglossaryentry{Tenor}
{
    name=tenor,
    description={Tenor refers to the amount of time left until the financial instrument expires. The maturity time points on a yield curve are also known as tenor points}
}

\newglossaryentry{bond}
{
    name=bond,
    description={A bond is a loan to a borrower made by a bondholder. The borrower (company, government, municipality) pays the loan principal of the bond to the bondholder on a maturity date as well as an interest for the life of the bond}
}

\newglossaryentry{coupon}
{
    name=coupon,
    description={A coupon is the annual interest rate paid on a bond, expressed as a percentage of the face value}
}

\newglossaryentry{Cbond}
{
    name=coupon bond,
    description={A bond on which interest is paid by coupons}
}

\newglossaryentry{arbitrage}
{
    name=arbitrage,
    description={Arbitrage is the simultaneous purchase and sale of an instrument to benefit from an imbalance in the price}
}

\newglossaryentry{Zbond}
{
    name=zero-coupon bond,
    description={A bond which does not pay interest and offers full face value profits at maturity}
}

\newglossaryentry{FR}
{
    name=forward rate,
    description={A forward rate is an interest rate applicable to the financial transaction that will take place in the future}
}

\newglossaryentry{option}{
    name=option,
    description={An option is a right, but not an obligation, to purchase or sell the underlying instrument at some time at a pre-defined price}
}

\newglossaryentry{call}{
    name=call,
    description={A call option gives a buyer the right, but not the obligation, to buy an instrument at a specified price}
}

\newglossaryentry{put}{
    name=put,
    description={A put option gives a buyer the right, but not the obligation, to sell an instrument at a specified price}
}

\newglossaryentry{strike}{
    name=strike price,
    description={A strike price is a fixed price at which an underlying instrument can be bought or sold}
}

\newglossaryentry{cap}{
    name=cap,
    description={An upper limit on the interest rate for a floating interest rate instrument}
}

\newglossaryentry{floor}{
    name=floor,
    description={An lower limit on the interest rate for a floating interest rate instrument}
}

\newglossaryentry{Euribor}{
    name=Euribor,
    description={Euribor is short for Euro Interbank Offered Rate. The Euribor rates are based on the average interest rates at which a large panel of European banks borrow funds from one another}
}

\newglossaryentry{return}{
    name=return,
    description={The amount of money returned to an account holder is typically referred to as the return}
}

\newglossaryentry{stock}{
    name=stock,
    description={Stock is a type of security that implies proportionate ownership in the issuing corporation}
}

\newglossaryentry{portfolio}{
    name=portfolio,
    description={In finance, a portfolio is a collection of investments held by an investment company, a hedge fund, a financial institution or an individual}
}

\newglossaryentry{ObsP}{
    name=observation period,
    description={One observation period is one working day at which the statistical data has been collected}
}

\newglossaryentry{RHP}{
    name=recommended holding period,
    description={The recommended holding period gives an idea to an investor that for how long should an investor hold the product to minimize the risk}
}

\newglossaryentry{Drift}{
    name=drift,
    description={The drift is the change of average value of a process}
}

\newglossaryentry{asset}{
    name=asset,
    description={An asset is something one owns, such as property, structures, money, or investments like stocks or bonds}
}

\newglossaryentry{RiskNeutral}{
    name=risk-neutral,
    description={Risk neutral term describes the investment which is insensitive to risk}
}

\newglossaryentry{Swap}{
    name=swaption,
    description={a. Swap: A swap is an agreement between two parties to exchange financial instruments for a certain time.\\
    b. Swaption: A swaption is an option to enter into a swap}
}

\begin{document}
%%%%%%%%%%%%%%%%%%%%%%%%%%%%%%%%%%%%%%%%%%%%%%%%%%%%%%%%%%%%%%%%%%%%%%%%%%%%%%%%%%
% First Page
%%%%%%%%%%%%%%%%%%%%%%%%%%%%%%%%%%%%%%%%%%%%%%%%%%%%%%%%%%%%%%%%%%%%%%%%%%%%%%%%%%
\hypersetup{pageanchor=false}
\graphicspath{{images/}}
%\def\keywords#1{{\bf Keywords}\\{#1}} %
%%%%%%%%%%%%%%%%%%%%%%%%%%%%%%%%%%%%%%%%%%%%%%%%%%%%%%%%%%%%%%%%%%%%%%%%%%%%%%%%%%
%% Change Title, Number and Version of Deliverable here!!
%%%%%%%%%%%%%%%%%%%%%%%%%%%%%%%%%%%%%%%%%%%%%%%%%%%%%%%%%%%%%%%%%%%%%%%%%%%%%%%%%%
\newcommand{\DelTitle}{Model order reduction for parametric high dimensional models in the analysis of financial risk}
\newcommand{\DelNumber}{D1}
\newcommand{\DelVersion}{1}
\newcommand{\footertext}{\raisebox{3mm}{Deliverable \DelNumber}}
\setlength{\footheight}{40pt}
\newcommand{\footerlogo}{\raisebox{3mm}{\leavevmode\includegraphics[width=1.5cm]{ROMSOC_Logo_bw}}}
\clearscrheadfoot
\pagestyle{empty}

\newcolumntype{C}{ >{\centering\arraybackslash} m{4cm} }

\begin{center}
Horizon 2020
\vspace{2cm}

  \begin{center}
  \includegraphics[width=0.2\textwidth]{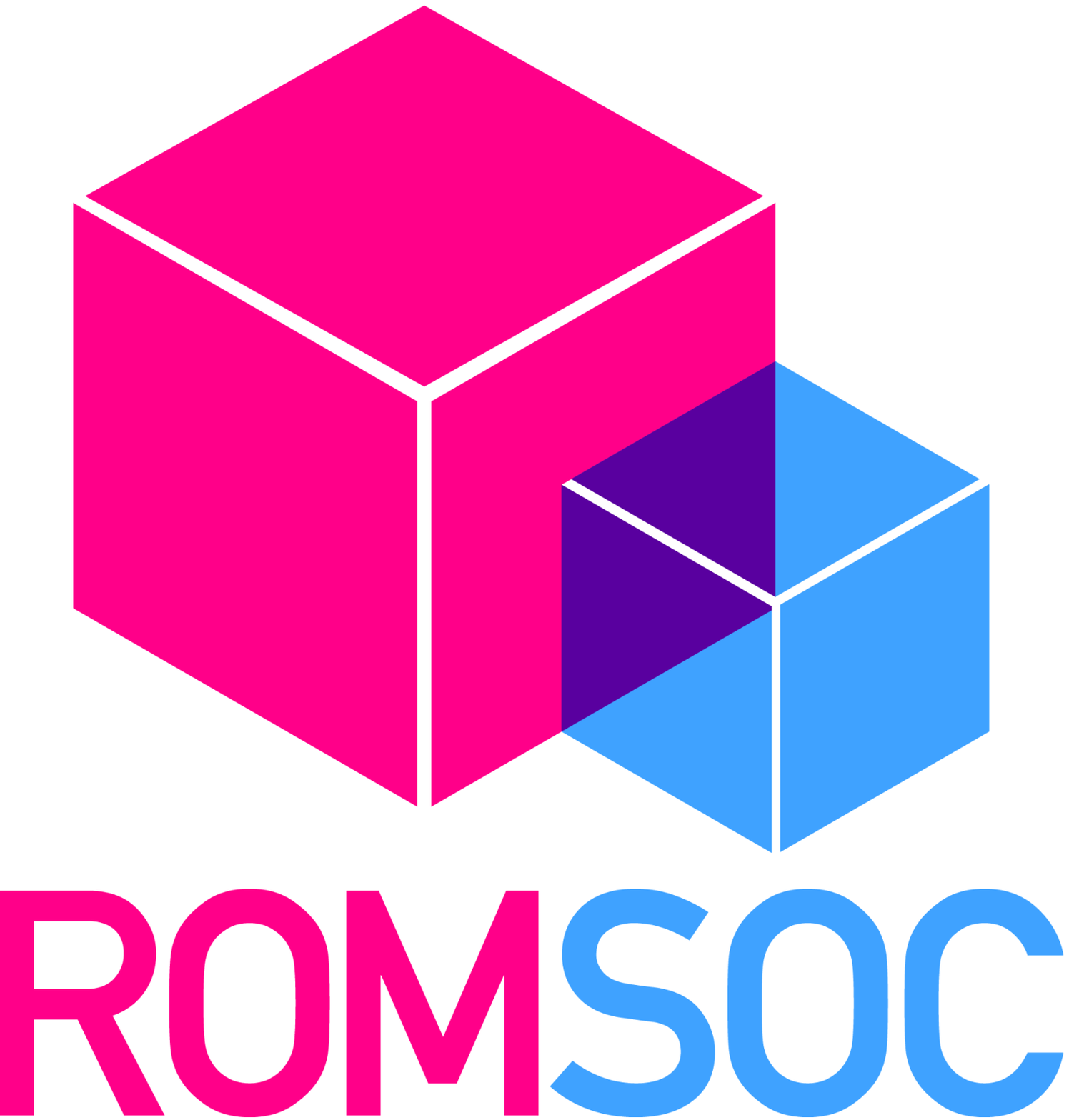}
  \vspace{2mm}
  \end{center}
  \vspace{1cm}
  {\Large Reduced Order Modelling, Simulation and Optimization of Coupled systems}
  \vspace{2cm}

  \begin{spacing}{2.5}
   \textbf{\Huge \DelTitle}\\\vspace{10mm}
   %5 \textbf{\Large Deliverable number: \DelNumber} \\\vspace{10mm} 
   % {\large Version \DelVersion}
  \end{spacing}
  
  \vspace*{\fill}

  %just to avoid warning
  \newcommand\undefcolumntype[1]{\expandafter\let\csname NC@find@#1\endcsname\relax}
  \newcommand\forcenewcolumntype[1]{\undefcolumntype{#1}\newcolumntype{#1}}
  \forcenewcolumntype{C}{ >{\arraybackslash} m{3cm} }

  \begin{tabular}{C@{\hspace*{0cm}}l}
    \includegraphics[scale=0.2]{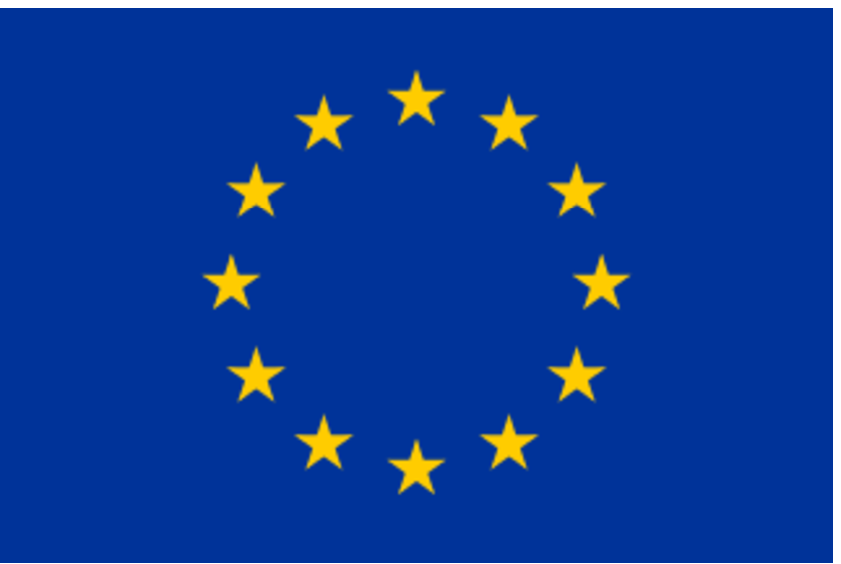} &
    \begin{tabular}{l}
    {European Union’s Horizon 2020 research and innovation programme}\\
    {under the Marie Sk\l odowska-Curie Grant Agreement No. 765374}\\
    \end{tabular}
  \end{tabular}
\end{center}

\clearpage

%%%%%%%%%%%%%%%%%%%%%%%%%%%%%%%%%%%%%%%%%%%%%%%%%%%%%%%%%%%%%%%%%%%%%%%%%%%%%%%%%%
% Second Page 
%%%%%%%%%%%%%%%%%%%%%%%%%%%%%%%%%%%%%%%%%%%%%%%%%%%%%%%%%%%%%%%%%%%%%%%%%%%%%%%%%%
\setlength{\headheight}{1cm}
\setlength{\footskip}{18mm}
\addtolength{\textheight}{-\footskip}
\pagestyle{empty}

\title{Model order reduction for parametric high dimensional interest rate models in the analysis of financial risk}
\date{}
% Authors
\author{Andreas Binder\footnotemark[1]
        \and Onkar Jadhav \footnotemark[2]
        \and Volker Mehrmann \footnotemark[3]}
\maketitle

\footnotetext[1]{MathConsult GmbH, Altenbergerstraße 69, A - 4040 Linz, Austria}
\footnotetext[2]{Institut f\"ur Mathematik MA 4-5, TU Berlin, Str. des 17. Juni 136, D-10623 Berlin, Germany}
\footnotetext[3]{Institut f\"ur Mathematik MA 4-5, TU Berlin, Str. des 17. Juni 136, D-10623 Berlin, Germany}
\section*{Abstract}
This paper presents a model order reduction (MOR) approach for high dimensional problems in the analysis of financial risk. 
To understand the financial risks and possible outcomes, we have to perform several thousand simulations of the underlying product.
These simulations are expensive and create a need for efficient computational performance. Thus, to tackle this problem, we establish a MOR approach based on a proper orthogonal decomposition (POD) method. The study involves the computations of high dimensional parametric convection-diffusion reaction partial differential equations (PDEs). POD requires to solve the high dimensional model at some parameter values to generate a reduced-order basis. We propose an adaptive greedy sampling technique based on surrogate modeling for the selection of the sample parameter set that is analyzed, implemented, and tested on the industrial data. The results obtained for the numerical example of a floater with a cap and floor under the Hull-White model indicate that the MOR approach works well for short-rate models.\\

\textbf{Keywords:} Financial risk analysis, short-rate models, convection-diffusion reaction equation, finite difference method, parametric model order reduction, proper orthogonal decomposition, adaptive greedy sampling, Packaged retail investment and insurance-based products (PRIIPs).\\

\textbf{MSC(2010):} 35L10, 65M06, 91G30, 91G60, 91G80
\vfill

\clearpage

\setcounter{tocdepth}{2}

\tableofcontents
%\vfill
%\listoffigures
%\begingroup
%\let\clearpage\relax
%\listoftables
%\vfill
%\endgroup

% This should be included if the deliverable becomes large
\newpage
%\input{acronyms} % optional

%\lofoot[\footerlogo \hspace{10pt} \footertext]{\footerlogo \hspace{10pt} \footertext}
%\lefoot[\footerlogo \hspace{10pt} \footertext]{\footerlogo \hspace{10pt} \footertext}

\refoot[\pagemark]{\pagemark}
\rofoot[\pagemark]{\pagemark}

\rohead{\rightmark}
\rehead{\rightmark}

\setcounter{page}{1}
\pagestyle{scrheadings}

% the following is optional and depend on the deliverable
\hypersetup{pageanchor=true}
%content.tex (optional)

\section{Introduction}
\emph{\gls{PRIIP}} (PRIIPs) are at the essence of the retail investment market. PRIIPs offer considerable benefits for retail investors which make up a market in Europe worth up to \euro 10 trillion. However, the product information provided by financial institutions to investors can be overly complicated and contains confusing legalese. To overcome these shortcomings, the EU has introduced new regulations on PRIIPs (European Parliament Regulation (EU) No 1286/2014) \cite{EUreg}. According to these regulations, a PRIIP manufacturer must provide a \emph{\gls{KID}} (KID) for an underlying product that is easy to read and understand. The KID informs about the vital features, such as costs and risks of the investment, before purchasing the product. The PRIIPs include interest rate derivatives such as the interest rate cap and floor \cite{Gupta05}, interest rate swaps \cite{Bicksler86}, etc.\\
A key information document includes a section about '\textit{what could an investor get in return?}' for the invested product which requires costly numerical simulations of financial instruments. This paper evaluates interest rate derivatives based on the dynamics of the short-rate models \cite{Brigo06}. For the simulations of short-rate models, techniques based on discretized convection-diffusion reaction partial differential equations (PDEs) are very successful \cite{Binder013}. To discretize the PDE, we implemented the finite difference method (FDM) \cite{Ekstrom09}. The FDM has been proven to be efficient for solving the short-rate models \cite{Haentjens12,Falcc13,Briani017}. The model parameters are usually calibrated based on market structures like \emph{\glspl{YC}}, \emph{\gls{cap}} volatilities, or \emph{\gls{Swap}} volatilities \cite{Brigo06}. The regulation demands to perform yield curve simulations for at least 10,000 times. A yield curve shows the interest rates varying with respect to 20-30 time points known as \emph{\gls{Tenor} points}. These time points are the contract lengths of an underlying instrument. The calibration based on several thousand simulated yield curves generates a high dimensional model parameter space as a function of these tenor points.
The 10,000 different simulated yield curves and the calibrated parameters based on these simulated yield curves can be considered as  10,000 different scenarios. We need to solve a high dimensional model (HDM) obtained by discretizing the short-rate PDE for such scenarios \cite{Cohen15}. Furthermore, the results obtained for these several thousand scenarios are used to calculate the possible values for an instrument under favorable, moderate, and unfavorable conditions. The favorable, moderate, and unfavorable scenario values are the values at 90th percentile, 50th percentile, and 10th percentile of 10,000 values, respectively. However, these evaluations are computationally costly, and additionally, have the disadvantage of being affected by the so-called \textit{curse of dimensionality} \cite{Piotr98}.\\
To avoid this problem, we establish a parametric model order reduction (MOR) approach based on a variant of the proper orthogonal decomposition (POD) method \cite{Chatterjee00,Berkooz93}. The method is also known as the Karhunen-Lo\'eve decomposition \cite{Graham96} or principal component analysis \cite{Jolliffe014} in statistics. 
The combination of a Galerkin projection approach and POD creates a powerful method for generating a reduced order model (ROM) from the high dimensional model that has a high dimensional parameter space \cite{Liang02}. This approach is computationally feasible as it always looks for low dimensional linear (or affine) subspaces \cite{Rathinam03,Vidal69}. Also, it is necessary to note that the POD approach considers the nonlinearities of the original system. Thus, the generated reduced order model will be nonlinear if the HDM is nonlinear as well.
POD generates an optimally ordered orthonormal basis in the least squares sense for a given set of computational data. Furthermore, the reduced order model is obtained by projecting a high dimensional system onto a low dimensional subspace obtained by truncating the optimal basis called reduced-order basis (ROB). The selection of the data set plays an important role and is most prominently obtained by the method of snapshots introduced in \cite{Sirovich87}. In this method, the optimal basis is computed based on a set of state solutions. These state solutions are known as snapshots and are calculated by solving the HDM for some pre-selected training parameter values. The quality of the ROM is bounded by the training parameters used to obtain the snapshots. Thus, it is necessary to address the question of how to generate the set of potential parameters which will create the optimal ROB. Some of the previous works implement either some form of fixed sampling or often only uniform sampling techniques \cite{Lucia04}. These approaches are straightforward, but they may neglect the vital regions in the case of high dimensional parameter spaces.\\
In the current work, a greedy sampling algorithm has been implemented to determine the best suitable parameter set \cite{Grepl05,Paul14,Amsallem015}. The basic idea is to select the parameters at which the error between the ROM and the HDM is maximal. Further, we compute the snapshots using these parameters and thus obtain the best suitable ROB which will generate a fairly accurate ROM. The calculation of the relative error between the ROM and the HDM is expensive, so instead, we use error estimators like the residual error associated with the ROM \cite{Thanh08,Veroy05}. The greedy sampling algorithm picks the optimal parameters which yield the highest values for the error estimator. Furthermore, we use these parameters to construct a snapshot matrix and, consequently, to obtain the desired ROB.\\
However, it is not reasonable to compute an error estimator for the entire parameter space. The error estimator is based on the norm of the residual, which scales with the size of the HDM. This problem forces us to select a pre-defined parameter set as a subset of the high dimensional parameter space to train the greedy sampling algorithm. We usually select this pre-defined subset randomly. But, a random selection may neglect the crucial parameters within the parameter space. Thus, to surmount this problem, we implemented an adaptive greedy sampling approach. We choose the most suitable parameters adaptively at each greedy iteration using an optimized search based on surrogate modeling. We construct a surrogate model for the error estimator and use it to find the best suitable parameters. The use of the surrogate model avoids the expensive computation of the error estimator over the entire parameter space. The adaptive greedy sampling approach associated with surrogate modeling has been introduced before in \cite{Paul14,Amsallem015}. There are several approaches to design a surrogate model like regression analysis techniques \cite{Lee015}, response surface models, or Kriging models \cite{jones01}.
The authors of \cite{Paul14} have designed a Kriging based surrogate model to select the most relevant parameters adaptively. However, in our case, due to the high dimensional parameter space, we may face the multicollinearity problem as some variable in the model can be written as a linear combination of the other variables in the model \cite{James013}. Also, we need to construct a surrogate model considering the fact that the model parameters are time-dependent. Thus, in this work, we construct a surrogate model based on the principal component regression (PCR) technique \cite{Lee015}. The PCR approach is a dimension reduction technique in which explanatory variables are replaced by few uncorrelated variables known as principal components. It replaces the multivariate problem with a more straightforward low dimensional problem and avoids overfitting.\\
In the classical greedy sampling approach, the convergence of the algorithm is observed using the error estimator. However, we can use the norm of the residual to estimate the exact error between the HDM and the ROM. In this work, we establish an error model for an exact error as a function of the error estimator based on the idea presented in \cite{Paul14}. Furthermore, we use this exact error model to observe the convergence of the greedy sampling algorithm.\\
To summarize, this paper presents an approach to select the most prominent parameters or scenarios for which we solve the HDM and obtain the required ROB. Thus, instead of performing 10,000 expensive computations, we perform very few expensive computations and solve the remaining scenarios with the help of the ROM. The paper illustrates the implementation of numerical algorithms and methods in detail. It is necessary to note that the choice of a short-rate model depends on the underlying financial instrument. In this work, we focus on one-factor short-rate models only. We implement the developed algorithms for the one-factor Hull-White model \cite{Hull90} and present the results with a numerical example of a floater with cap and floor \cite{Fabozzi98}. The current research findings indicate that the MOR approach works well for short-rate models.\\
The paper is organized as follows. Section \ref{MH} presents a model hierarchy to construct a short-rate model along with some of the well-known one-factor models. Section \ref{YCnPC} is branched into two parts. In subsection \ref{YCS}, we present a detailed simulation procedure for yield curves and, subsequently, subsection \ref{ParaCalb} describes the calibration of model parameters based on these simulated yield curves. In section \ref{NumericalMethod}, subsection \ref{FDM} illustrates the FDM implemented for the Hull-White model, and subsection \ref{MORPOD} introduces the projection-based MOR technique for the HDM. The selection of best suitable parameters to obtain the ROB based on the classical greedy approach is presented in subsection \ref{CG}. To overcome the drawbacks associated with the classical greedy approach, subsection \ref{AG} explains the adaptive greedy method with the surrogate modeling technique illustrated in sub-subsection \ref{SM}. Furthermore, sub-subsection \ref{AGAlgo} presents a detailed algorithm and its description for the adaptive greedy approach. Finally, we tested our algorithms using a numerical example of a floater, and the obtained results are presented in section \ref{NEx}.
\section{Model Hierarchy}
\label{MH}
The management of interest rate risks, i.e., the control of change in future cash flows due to the fluctuations in interest rates is of great importance. Especially, the pricing of products based on the stochastic nature of the interest rate creates the necessity for mathematical models.
In this section, we present some basic definitions and the model hierarchy to construct a financial model for any underlying instrument. 
\subsection{Bank Account and Short-Rate}
First we introduce the definition for a \emph{\gls{bank}} also called as a money-market account. When investing a certain amount in a bank account, we expect it to grow at some rate over time. A money-market account represents a risk-less investment with a continuous profit at a risk-free rate.
\theoremstyle{definition}
\begin{definition}{\textbf{Bank account (Money-market account)}.}
\textit{Let $B(t)$ be the value of a bank account at time $t \geq 0$. We assume that the bank account evolves according to the following differential equation with $B(0)=1$},
\begin{equation}
    dB(t) = B(t) r_t dt,
    \label{1}
\end{equation}
where $r_t$ is a \emph{\gls{shortrate}}. This gives
\begin{equation}
    B(t) = \mathrm{exp} \bigg ( \int_{o}^{t} r_s ds \bigg ).
    \label{2}
\end{equation}
\end{definition}
According to the above definition, investing a unit amount at time $t=0$ yields the value in (\ref{2}) at time $t$, and $r_t$ is the short-rate at which the bank account grows. This rate $r_t$ is the growth rate of the bank account $B$ within a small time interval $(t,t+dt)$. 
When working with interest-rate products, the study about the variability of interest rates is essential. Therefore, it is necessary to consider the short-rate as a stochastic process and it prompts us to construct a stochastic model that will describe the dynamics of the short-rate.\\
Let $S$ be the price of a \emph{\gls{stock}} at the end of the $n$th trading day. The \emph{daily \gls{return}} from days $n$ to $(n+1)$ is given by $(S_{n+1}-S_n)/S_n$. In general, it is common to work with log returns, since the log return of $k$ days can be easily computed by adding up the daily log returns:
\begin{equation*}
    \mathrm{log}(S_k/S_0) = \mathrm{log}(S_1/S_0) + \cdots + \mathrm{log}(S_k/S_{k-1}).
\end{equation*}
Based on the assumption that the log returns over disjoint time intervals are stochastically independent, and are equally distributed, the central limit theorem \cite{Dudley10} of probability theory implies that the log returns are normally distributed \cite{Corhay94}.
So, it is necessary to define a stochastic model in continuous time in which log returns over arbitrary time intervals are normally distributed. The \emph{Brownian motion} provides these properties \cite{Albrecher13}.
\theoremstyle{definition}
\begin{definition}{\textbf{Brownian motion.}}
\textit{A standard Brownian motion is a stochastic process $W(t)$ where$ \, \, t \in \mathbb{R}$, i.e., a family of random variables $W(t)$, indexed by non-negative real numbers $t$ with the following properties:}
\begin{itemize}
    \item \textit{At} $t=0$, $W_0=0$.
    \item \textit{With probability 1, the function $W(t)$ is continuous in $t$.}
    \item \textit{For $t\geq 0$, the increment $W(t+s)-W(s)$ is normally distributed with mean 0 and variance $t$, i.e.,}
    \begin{equation*}
        W(t+s) - W(s) \sim N(0,t).
    \end{equation*}
    \item \textit{For all $n$ and times $t_0 < t_1 < \cdots < t_{n-1} < t_n$, the increments $W({t_j})-W({t_{j-1}})$ are stochastically independent.}
\end{itemize}
\end{definition}
Based on the definition of the Brownian motion, we can establish a stochastic differential equation (SDE).
Consider an ordinary differential equation (ODE)
\begin{equation}
    \frac{dX(t)}{dt} = q(t)X(t),
    \label{3}
\end{equation}
with an initial condition $X(0) = X_0$.
When we consider ODE (\ref{3}) with an assumption that the parameter $q(t)$ is not a deterministic but rather a stochastic parameter, we get a stochastic differential equation.\\
In our case, the stochastic parameter $q(t)$ is given as \cite{Grigoriu02}
\begin{equation*}
    q(t) = f(t) + h(t)w (t),
\end{equation*}
where ${w} (t)$ is a white noise process.
Thus, we get
\begin{equation}
    \frac{dX(t)}{dt} = f(t)X(t) + h(t)X(t)w (t).
    \label{4}
\end{equation}
This equation is known as Langevin equation \cite{Mahnke08}. Here $X(t)$ is a stochastic variable having the initial condition $X(0) = X_0$ with probability one. The Langevin force $w(t)=dW(t)/dt$ is a fluctuating quantity having Gaussian distribution.
Substituting $dW(t) = w(t) dt$ in (\ref{4}), we get
\begin{equation}
    dX(t) = f(t)X(t)dt + h(t)X(t)dW(t)
    \label{5}
\end{equation}
In the general form a stochastic differential equation is given by
\begin{equation}
    dX(t) = f(t,X(t))dt + g(t,X(t))dW(t),
    \label{6}
\end{equation}
where $f(t,X(t)) \in \mathbb{R}$, and $g(t,X(t)) \in \mathbb{R}$ are sufficiently smooth functions.
Based on (\ref{6}), we can define a stochastic differential equation with the short-rate $r_t$ as a stochastic variable as
\begin{equation}
    dr_t = f(t,r_t)dt + g(t,r_t)dW(t),
    \label{7}
\end{equation}
Furthermore, based on Ito's lemma \cite{Wilmott02}, we can derive a general PDE for any underlying instrument depending on the short-rate $r$.
\begin{theorem}{\textbf{Ito's Lemma.}}
\textit{Let $\xi(X(t),t)$ be a sufficiently smooth function and let the stochastic process $X(t)$ be given by (\ref{8}), then with probability one,}
\begin{equation}
\begin{aligned}
    d\xi(X(t),t) &= \bigg(\frac{\partial \xi}{\partial X(t)}f(X(t),t) + \frac{\partial \xi}{\partial t} + \frac{1}{2} \frac{\partial^2 \xi}{\partial X^2(t)} g^2(X(t),t) \bigg) dt \\
    &+ \frac{\partial \xi}{\partial X(t)} g(X(t),t)dW(t).
    \end{aligned}
    \label{8}
\end{equation}
\end{theorem}
Consider a \emph{\gls{RiskNeutral}} \emph{\gls{portfolio}} $\Pi_t$ that depends on the short-rate $r_t$ and consists of two interest rate instruments $V_1$ and $V_2$ with different maturities $T_1$ and $T_2$ respectively. Let there be $\Delta$ units of the instrument $V_2$. For an infinitesimal time interval, the value change of the portfolio is $d\Pi_t = \Delta dV_2 - dV_1$. To avoid the \emph{\gls{arbitrage}}, we have to consider a risk-free rate \cite{Binder013}, which gives
\begin{equation*}
    d\Pi_t = \Delta dV_2  + (V_1 - \Delta V_2)rdt - dV_1.
\end{equation*}
Based on Ito's lemma, we get
\begin{equation*}
    \begin{aligned}
    d\Pi_t &= (V_1 - \Delta V_2)r_tdt\\ 
    &- \bigg[ \bigg(\frac{\partial V_1}{\partial r_t}f(r_t,t) + \frac{\partial V_1}{\partial t} + \frac{1}{2} \frac{\partial^2 V_1}{\partial r^2_t} g^2(r_t,t) \bigg) dt + \frac{\partial V_1}{\partial r_t} g(r_t,t)dW(t) \bigg] \\
    &+ \Delta \bigg[ \bigg(\frac{\partial V_2}{\partial r_t}f(r_t,t) + \frac{\partial V_2}{\partial t} + \frac{1}{2} \frac{\partial^2 V_2}{\partial r^2_t} g^2(r_t,t) \bigg) dt 
    + \frac{\partial V_2}{\partial r_t} g(r_t,t)dW(t) \bigg].
    \end{aligned}
\end{equation*}
Let $\Delta = \bigg (\frac{\partial V_1}{\partial r_t} \bigg/\frac{\partial V_2}{\partial r_t}\bigg )$. Due to the zero net investment requirement, i.e., $d\Pi_t = 0$, we obtain
\begin{equation*}
    \begin{aligned}
    0 &= \bigg [V_1 - \bigg (\frac{\partial V_1}{\partial r_t} \bigg/\frac{\partial V_2}{\partial r_t}\bigg ) V_2\bigg ] r_tdt\\
    &- \bigg[ \bigg(\frac{\partial V_1}{\partial r_t}f(r_t,t) + \frac{\partial V_1}{\partial t} + \frac{1}{2} \frac{\partial^2 V_1}{\partial r^2_t} g^2(r_t,t) \bigg) dt + \frac{\partial V_1}{\partial r_t} g(r_t,t)dW(t) \bigg] \\
    &+ \bigg (\frac{\partial V_1}{\partial r_t} \bigg/\frac{\partial V_2}{\partial r_t} \bigg ) \bigg[ \bigg(\frac{\partial V_2}{\partial r_t}f(r_t,t) + \frac{\partial V_2}{\partial t} + \frac{1}{2} \frac{\partial^2 V_2}{\partial r^2_t} g^2(r_t,t) \bigg) dt 
    + \frac{\partial V_2}{\partial r_t} g(r_t,t)dW(t) \bigg].
    \end{aligned}
\end{equation*}
Eliminating the stochastic term, we get
\begin{equation*}
    \begin{aligned}
    &\bigg [V_1 - \bigg (\frac{\partial V_1}{\partial r_t} \bigg/\frac{\partial V_2}{\partial r_t}\bigg ) V_2\bigg ] r_tdt\\ 
    &= \bigg[ \frac{\partial V_1}{\partial t} + \frac{1}{2} \frac{\partial^2 V_1}{\partial r^2_t} g^2(r_t,t) - \bigg (\frac{\partial V_1}{\partial r_t} \bigg/\frac{\partial V_2}{\partial r_t} \bigg ) \bigg (\frac{\partial V_2}{\partial t} + \frac{1}{2} \frac{\partial^2 V_2}{\partial r^2_t} g^2(r_t,t) \bigg) \bigg ]dt.
    \end{aligned}
\end{equation*}
Rearranging the terms, and using the notation $r=r_t$, we get
\begin{equation*}
    \frac{\frac{\partial V_1}{\partial t} + \frac{1}{2} \frac{\partial^2 V_1}{\partial r^2} g^2(r,t) - rV_1}{\frac{\partial V_1}{\partial r}} =  \frac{\frac{\partial V_2}{\partial t} + \frac{1}{2} \frac{\partial^2 V_2}{\partial r^2} g^2(r,t) - rV_2}{\frac{\partial V_2}{\partial r}}.
\end{equation*}
Denoting the right side as $u(r,t)$
\begin{equation*}
    \frac{\frac{\partial V_1}{\partial t} + \frac{1}{2} \frac{\partial^2 V_1}{\partial r^2} g^2(r,t) - rV_1}{\frac{\partial V_1}{\partial r}} =  u(r,t),
\end{equation*}
and using the notation $V = V_1$, we obtain the following PDE for any financial instrument $V$ depending on $r$
\begin{equation}
    \frac{\partial V}{\partial t} + \frac{1}{2} g^2(r,t) \frac{\partial^2 V}{\partial r^2} - u(r,t) \frac{\partial V}{\partial r} - rV = 0,
    \label{9}
\end{equation}
We introduce some well-known one-factor short-rate models in the following subsections.
\subsection{Vasicek and Cox-Ingersoll-Ross Models}
\label{VnCIR}
The following SDE represent two different models depending on the choice of a parameter $\lambda$.
\begin{equation}
    dr_t = (a - br_t)dt + \sigma r^\lambda dW(t),
    \label{10}
\end{equation}
where $a$, $b$, $\sigma$, and $\lambda$ are positive constants. The model proposed in \cite{Vasicek77} considers $\lambda = 0$ and is well known as the Vasicek model, while the Cox-Ingersoll-Ross model introduced in \cite{CIR85} considers $\lambda = 0.5$.
One of the drawbacks of the Vasicek model is that the short-rate can be negative. On the other hand, in the case of the Cox-Ingersoll-Ross model, the square root term does not allow negative interest rates.
However, the major drawback of these models is that the model parameters are constant, so we can not fit the models to the market structures like yield curves.
\subsection{Hull-White Model}
\label{HWM}
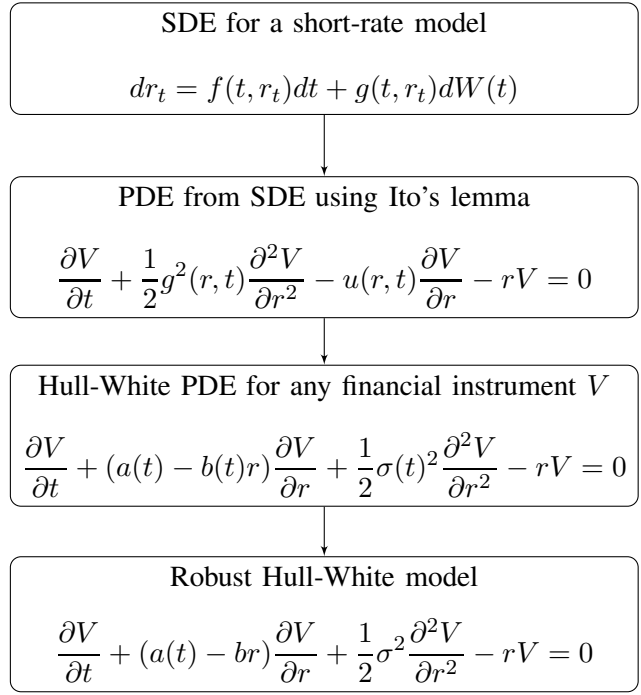
\begin{figure}[H]
\begin{center}
\tikzstyle{block} = [rectangle, draw,
    text width=8cm, text centered, rounded corners, minimum height=3em]
\tikzstyle{line} = [draw, -latex']

\begin{tikzpicture}[node distance = 2.5cm, auto]
\node [block] (MainModel) {SDE for a short-rate model
\begin{equation*}
    dr_t = f(t,r_t)dt + g(t,r_t)dW(t)
\end{equation*}
};
\node[block, below of=MainModel] (PDE){PDE from SDE using Ito's lemma
\begin{equation*}
        \frac{\partial V}{\partial t} + \frac{1}{2} g^2(r,t) \frac{\partial^2 V}{\partial r^2} - u(r,t) \frac{\partial V}{\partial r} - rV = 0
\end{equation*}
};
\node[block, below of=PDE](HWPDE){Hull-White PDE for any financial instrument $V$
\begin{equation*}
    \frac{\partial V}{\partial t} + (a(t) - b(t)r)\frac{\partial V}{\partial r} + \frac{1}{2} \sigma(t)^2 \frac{\partial^2 V}{\partial r^2} - rV = 0
\end{equation*} 
};
\node[block, below of=HWPDE](HWPDERobust){Robust Hull-White model
\begin{equation*}
    \frac{\partial V}{\partial t} + (a(t) - br)\frac{\partial V}{\partial r} + \frac{1}{2} \sigma^2 \frac{\partial^2 V}{\partial r^2} - rV = 0
\end{equation*} 
};
\path [line] (MainModel) -- (PDE);
\path [line] (PDE) -- (HWPDE);
\path [line] (HWPDE) -- (HWPDERobust);
\end{tikzpicture}
\end{center}
\caption{A model hierarchy to construct the Hull-White model based on the short-rate $r$ with constant $b$ and $\sigma$.}
\label{fig:1}
\end{figure}

The Hull-White model \cite{Hull90,Hull93} is an extension of the Vasicek model which can be fitted to market structures like yield curves. The SDE is given as
\begin{equation}
    dr_t = (a(t) - b(t)r_t)dt + \sigma(t)dW(t), 
    \label{11}
\end{equation}
where now the parameters $a(t)$, $b(t)$, and $\sigma(t)$ are time-dependent parameters. The term $(a(t) - b(t)r_t)$ is a drift term and $a(t)$ is known as \emph{deterministic \gls{Drift}}.
We can define a PDE for any underlying instrument based on the Hull-White model depending on $r$. In the case of the Hull-White model in (\ref{9}), $g(r,t) = \sigma(t)$, and $-u(r,t) = (a(t) - b(t)r)$. Substituting $g(r,t)$ and $u(r,t)$, we get
\begin{equation}
    \frac{\partial V}{\partial t} + (a(t) - b(t)r) \frac{\partial V}{\partial r} + \frac{1}{2} \sigma^2(t) \frac{\partial^2 V}{\partial r^2} - rV = 0,
    \label{12}
\end{equation}
In the following, we consider the parameter $b(t)$ and $\sigma(t)$ as constants to have a robust model.
Hull and White supported for making model parameters $b$ and $\sigma$ to be independent of time \cite{Hull94}. The problem is that the volatility term structure in the future could become nonstationary in the sense that the future term structure implied by the model can be quite different than it is today. Also, the author of \cite{Darbellay98} quoted that the volatility in the future may reach zero, which ultimately could result in implausible option prices, thus, we suggest to consider parameters $b$ and $\sigma$ as independent of time. Figure \ref{fig:1} shows the model hierarchy for the Hull-White model.
The following section \ref{YCnPC} presents the simulation procedure for yield curves and the calibration of the model parameter $a(t)$ based on these simulated yield curves.
\section{Yield Curve Simulation and Parameter Calibration}
\label{YCnPC}
\subsection{Yield Curve Simulation}
\label{YCS}
The problem of determining the model parameters is relatively complex. The time-dependent parameter $a(t)$ is derived from yield curves, which determine the average direction in which the short-rate $r$ moves. The PRIIP regulation demands to perform yield curve simulations for at least 10,000 times \cite{EUreg}. We explain the detailed yield curve simulation procedure in this subsection.
\begin{enumerate}
    \item Collect historical data for the interest rates.
\end{enumerate}
The data set must contain at least 2 years of daily interest rates for an underlying instrument or 4 years of weekly interest rates or 5 years of monthly interest rates. Further, we construct a data matrix $D \in \mathbb{R}^{n\times m}$ of the collected historical interest rates data, where each row of the matrix forms a yield curve, and each column is a tenor point $m$. The tenor points are the different contract lengths of an underlying instrument. For example, we have collected the daily interest rate data at 20-30 tenor points in time over the past five years. Each year has approximately 260 working days also known as \emph{\glspl{ObsP}}. Thus, there are $n \approx 1306$ observation periods and $m \approx 20$ tenor points in time.
\begin{enumerate}
\setcounter{enumi}{1}
    \item Calculate the log return over each period.
\end{enumerate}
We take the natural logarithm of the ratio between the interest rate at each observation period and the interest rate at the preceding period. To avoid problems while taking the natural logarithm, we have to ensure that all elements of the data matrix $D$ are positive which is done by adding a correction term $\gamma$.
 \begin{equation*}
 \begin{aligned}
     \Bar{D} &= D + \gamma W,\\
     \Bar{d}_{ij} &= d_{ij} + \gamma w_{ij}, \hspace{2cm} w_{ij}=1 \, \, \mathrm{for}\, \mathrm{all} \,\, i,j
     \end{aligned}
 \end{equation*}
The correction factor $\gamma$ ensuring all elements of matrix $D$ to be positive. Here the matrix $W \in \mathbb{R}^{n\times m}$ is a binary matrix having all entries as 1. 
The selection of $\gamma$ does not affect the final simulated yield curves as we are compensating this shift at the bootstrapping stage by subtracting it from the simulated rates.
Then we calculate the log returns over each period and store them into a new matrix $\hat{D} = \hat{d}_{ij} \in \mathbb{R}^{n\times m}$ as
\begin{equation*}
    \hat{d}_{ij} = \frac{\mathrm{ln}(\Bar{d}_{ij})}{\mathrm{ln}(\Bar{d}_{(i-1)j})}.
\end{equation*}
\begin{enumerate}
\setcounter{enumi}{2}
    \item Correct the returns observed at each tenor so that the resulting set of returns at each tenor point has a zero mean.
\end{enumerate}
We calculate the arithmetic mean $\mu_j$ of each column of the matrix $\hat{D}$,
\begin{equation*}
    \mu_j = \frac{1}{n} \sum_{i=1}^{n} \hat{d}_{ij},
\end{equation*}
and subtract this arithmetic mean $\mu_j$ from each element of the corresponding $j$th column of a matrix $\hat{D}$ and store the obtained results in the matrix $\Bar{\Bar{D}} \in \mathbb{R}^{n\times m}$,
\begin{equation*}
    \Bar{\Bar{d}}_{ij} = \hat{d}_{ij} - \mu_j w_{ij}.
\end{equation*}
\begin{enumerate}
\setcounter{enumi}{3}
    \item Compute the singular value decomposition \cite{Golub70} of the matrix $\Bar{\Bar{D}}$.
\end{enumerate}
The singular value decomposition of the matrix $\Bar{\Bar{D}}$ is
\begin{equation*}
    \Bar{\Bar{D}} = \Phi \Sigma \Psi^T,
\end{equation*}
\begin{equation*}
\Bar{\Bar{D}}= 
\begin{bmatrix}
\phi_{11} & \cdots & \phi_{1m}\\ 
\vdots & \vdots & \vdots\\  
\phi_{m1} & \cdots & \phi_{mm}\\
\end{bmatrix}_{m \times m}
\cdot
\begin{bmatrix}
\Sigma_{1} & 0 & \cdots\\ 
0 & \ddots & \vdots\\ 
\vdots & \cdots & \Sigma_{m}\\
\end{bmatrix}_{m \times m}
\cdot
\begin{bmatrix}
\psi_{11} & \cdots & \psi_{1m}\\
\vdots & \vdots & \vdots\\
\psi_{m1} & \cdots & \psi_{mm}\\
\end{bmatrix}_{m \times m}
\end{equation*}
where $\Sigma$ is a diagonal matrix having singular values $\Sigma_i$ arranged in descending order.
The columns of $\Phi$ are the normalized singular vectors $\phi \in \Phi$, and the columns of $\Phi \Sigma$ are known as principal components. The right singular vectors $\psi \in \Psi$ are the eigenvectors or also called \emph{principal directions} of the covariance matrix $\mathcal{C} = \Bar{\Bar{D}}^T\Bar{\Bar{D}}$. A detailed relation between the singular value decomposition and the principal component analysis is presented in the Appendix \ref{SVDnPCA}.
%\clearpage
\begin{enumerate}
\setcounter{enumi}{4}
    \item Select the principal singular vectors corresponding to the maximum energy.
\end{enumerate}
The relative importance of $i$th principal singular-vector is determined by the relative energy $\Xi_i$ of that component defined as
\begin{equation*}
    \Xi_i = \frac{\Sigma_i}{\sum_{i=1}^{m} \Sigma_i},
\end{equation*}
where the total energy is given by $\sum_{i=1}^{m} \Xi_i =1$. We select $p$ right singular vectors corresponding to the maximum energies from the matrix $\Psi$. We construct a matrix $\Bar{\Psi} \in \mathbb{R}^{m\times p}$ composed of these selected singular vectors
\begin{equation*}
    \Bar{\Psi} = 
\begin{bmatrix}
\psi_{11} & \cdots & \psi_{1p}\\ 
\vdots & \vdots & \vdots\\  
\psi_{m1} & \cdots & \psi_{mp}\\
\end{bmatrix}_{m \times p}
\end{equation*}
\begin{enumerate}
\setcounter{enumi}{5}
    \item Calculate the matrix of returns to be used for the simulation of yield curves.
\end{enumerate}
We project the matrix $\Bar{\Bar{D}}$ onto the matrix of selected singular vectors $\Bar{\Phi}$.
\begin{equation*}
    \mathfrak{X} = \Bar{\Bar{D}} \cdot \Bar{\Psi}, \hspace{2cm} \mathfrak{X} \in \mathbb{R}^{n\times p}.
\end{equation*}
Furthermore, we calculate the matrix of returns $M_R \in \mathbb{R}^{n\times m}$ by multiplying the matrix $\mathfrak{X}$ with the transpose of the matrix of singular vectors $\Bar{\Psi}$.
\begin{equation}
    M_R = \mathfrak{X} \cdot \Bar{\Psi}^T.
    \label{13}
\end{equation}
This process simplifies the statistical data $\Bar{\Bar{D}}$ that transforms $m$ correlated tenor points into $p$ uncorrelated principal components. It allows reproducing the same data by simply reducing the total size of the model.
\begin{enumerate}
\setcounter{enumi}{6}
    \item Bootstrapping
\end{enumerate}
Bootstrapping is a type of resampling where large numbers of small samples of the same size are drawn repeatedly from the original data set.
It follows the law of large numbers, which states that if samples are drawn over and over again, then the resulting set should resemble the actual data set. In short, the bootstrapping creates different scenarios based on simulated samples, which resembles the actual data. These scenarios can be further used to obtain the values at favorable, moderate, and unfavorable conditions for an underlying instrument.
According to the PRIIP regulations, we have to perform a bootstrapping procedure for the yield curve simulation for at least 10,000 times. The regulations state that a standardized key information document shall include the minimum \emph{\gls{RHP}} (RHP).
\theoremstyle{definition}
\begin{definition}{\textbf{Holding period.}}
\textit{A holding period is a period between the acquisition of an asset and its sale. It is the length of time during which an underlying instrument is held by an investor.}
\end{definition}
\begin{remark}
The recommended holding period gives an idea to an investor for how long should an investor hold the product to minimize the risk. Generally, the RHP is given in years.
\end{remark}
The time step in the simulation of yield curves is typically one observation period. Let $h$ be the RHP in days (e.g., $h \approx$ 2600 days or 10 years). So, there are $h$ observation periods in the RHP.
For each observation period in the RHP, we select a row at random from the matrix $M_R$, i.e., we select $h$ random rows from the matrix $M_R$. We construct a matrix $\Bar{M_R} = \mathfrak{\chi}_{ij} \in \mathbb{R}^{h\times m}$ from these selected random rows.
Further, we sum over the selected rows of the columns corresponding to the tenor point $j$,
\begin{equation*}
    \Bar{\chi}_j = \sum_{i=1}^{h} \mathfrak{\chi}_{ij}, \hspace{2cm} j = 1, \cdots, m.
\end{equation*}
In this way, we obtain a row vector $\Bar{\chi}\in \mathbb{R}^{1\times m}$ such that
\begin{equation*}
    \Bar{\chi} = [\Bar{\chi}_1 \, \, \Bar{\chi}_2 \, \, \cdots \, \, \Bar{\chi}_m].
\end{equation*}
The final simulated yield rate $y_j$ at tenor point $j$ is the rate at the last observation period $\Bar{d}_{nj}$ at the corresponding tenor point $j$,
\begin{enumerate}
    \item multiplied by the exponential of the $\Bar{\chi}_j$,
    \item adjusted for any shift $\gamma$ used to ensure positive values for all tenor points.
    \item adjusted for the forward rate so that the expected mean matches current expectations.
\end{enumerate}
The forward rate between two time points $t_1$ and $t_2$ is then given as
\begin{equation*}
    r_{1,2} = \frac{r(t_0,t_2)(t_2 - t_0) - r(t_0,t_1)(t_1 - t_0)}{t_2 - t_1},
\end{equation*}
where $t_1$ and $t_2$ are measured in years. Here, $r(t_0,t_1)$ and $r(t_0,t_1)$ are the interest rates available from the data matrix for the time periods $(t_0,t_1)$ and $(t_0,t_2)$.
Thus, the final simulated rate $r_j$ is given by
\begin{equation}
    y_j = \Bar{d}_{nj} \times \mathrm{exp}({\Bar{\chi}_j}) - \gamma w_{nj} + r_{1,2}, \hspace{2cm} j = 1,\cdots,m.
    \label{14}
\end{equation}
Finally, we get the simulated yield curve from the calculated simulated returns $y_j$ as
\begin{equation*}
    y = [y_1 \, \, y_2 \, \, \cdots \, \, y_m], \hspace{2cm} j = 1,\cdots,m.
\end{equation*}
We perform the bootstrapping procedure for at least $s = 10,000$ times and construct a simulated yield curve matrix $Y \in \mathbb{R}^{s\times m}$ as
\begin{equation}
    Y = 
\begin{bmatrix}
y_{11} & \cdots & y_{1m}\\ 
\vdots & \vdots & \vdots\\  
y_{s1} & \cdots & y_{sm}\\
\end{bmatrix}_{s \times m}
\label{15}
\end{equation}
Subsection \ref{ParaCalb} explains the parameter calibration based on simulated yield curves.
\subsection{Parameter Calibration}
\label{ParaCalb}
The model parameters $a(t)$ is calibrated based on simulated yield curves $Y$.
According to \cite{Shreve04}, we can write a closed-form solution for a zero-coupon bond $B(t,T)$ maturing at time $T$ based on the Hull-White model as
\begin{equation}
    B(t,T) = \mathrm{exp}\{-r(t)\Gamma(t,T) - \Lambda(t,T)\},
    \label{16}
\end{equation}
where $r$ is the short rate at time $t$ and 
\begin{equation}
\begin{aligned}
        \kappa(t) &= \int_0^t b(s)ds,\\
        \Gamma(t,T) &= \int_t^T e^{-\kappa(t)} dt,\\
        \Lambda(t,T) &= \int_t^T \bigg [ e^{\kappa(v)}a(v) \bigg ( \int_v^T e^{-\kappa(z)}dz \bigg ) - \frac{1}{2} e^{2\kappa (v)} \sigma^2(v) \bigg ( \int_v^T e^{-\kappa(z)} dz \bigg)^2 \bigg ] dv.
\end{aligned}
\label{17}
\end{equation}
We take the following input data for the calibration:
\begin{enumerate}
    \item The zero-coupon bond prices for all maturities $T_m$, $0 \leq T_m \leq T$, where $T_m$ is the maturity at the $m$th tenor point.
    \item The initial value of $a(t)$ at $t=0$ as $a(0)$.
    \item The constant value of the volatility $\sigma$ of the short-rate $r_t$ at all maturities $0 \leq T_m \leq T$ is assumed to be constant.
    \item The constant value of the parameter $b$ is known and constant for all maturities $0 \leq T_m \leq T$. 
\end{enumerate}
We then determine the value of $\Gamma(0,T)$ as follows:
\begin{equation}
\begin{aligned}
    e^{-\kappa(T)} &= \frac{\partial }{\partial T} \Gamma(0,T),\\
    \kappa(T) &= -\mathrm{ln} \frac{\partial}{\partial T} \Gamma(0,T),\\
    \frac{\partial}{\partial T} \kappa(T) &= \frac{\partial}{\partial T} \int_0^T b(s) ds = b(T).
    \label{18}
\end{aligned}
\end{equation}
As we know the value of $b$ from the given initial condition, we can compute $\kappa(t)$. Subsequently, using $\kappa(t)$, we compute $\Gamma(t)$.\\
We can use $\Lambda(0,T)$ to determine $a(t)$, for $0 \leq T_m \leq T$ in the following way:
\begin{equation}
\begin{aligned}
\frac{\partial}{\partial T} \Lambda(0,T) &= \int_{0}^{T} \Bigg [e^{\kappa(v)} a(v) e^{-\kappa(T)}\\
&- e^{2\kappa(v)}\sigma^2(v)e^{-\kappa(T)} \bigg ( \int_{v}^{T} e^{-\kappa(z)} dz \bigg ) \Bigg ] dv, \\
e^{\kappa(T)} \frac{\partial}{\partial T} \Lambda(0,T) &= \int_{0}^{T} \Bigg [e^{\kappa(v)} a(v) - e^{2\kappa(v)}\sigma^2(v) \bigg ( \int_{v}^{T} e^{-\kappa(z)} dz \bigg ) \Bigg ] dv,\\
\frac{\partial}{\partial T} \Bigg [ e^{\kappa(T)} \frac{\partial}{\partial T} \Lambda(0,T) \Bigg ] &= e^{\kappa(T)} a(T) - \int_{0}^{T} e^{2\kappa(v)}\sigma^2(v)e^{-\kappa(T)} dv,\\
e^{\kappa(T)} \Bigg [ e^{\kappa(T)} \frac{\partial}{\partial T} \Lambda(0,T) \Bigg ] &= e^{2\kappa(T)} a(T) - \int_{0}^{T} e^{2\kappa(v)}\sigma^2(v) dv,\\
\frac{\partial}{\partial T} \Bigg [ e^{\kappa(T)} \Bigg [ e^{\kappa(T)} \frac{\partial}{\partial T} \Lambda(0,T) \Bigg ]\Bigg ] &= \frac{\partial a(T)}{\partial T} e^{2\kappa(T)} + 2a(T) e^{2\kappa(T)} \frac{\partial}{\partial T}\kappa(T) - e^{2\kappa(T)}\sigma^2(T),\\
 \frac{\partial}{\partial T} \Bigg [ e^{\kappa(T)} \Bigg [ e^{\kappa(T)} \frac{\partial}{\partial T} \Lambda(0,T) \Bigg ]\Bigg ] &= \frac{\partial a(T)}{\partial T} e^{2\kappa(T)} + 2a(T)e^{2\kappa(T)}b(T) - e^{2\kappa(T)}\sigma^2(T).   
\end{aligned}
\label{19}
\end{equation}
The yield $y(T)$ at time $T$ is given by
\begin{equation}
    y(T) = -\mathrm{ln}B(0,T).
    \label{20}
\end{equation}
From (\ref{16}), we can obtain
\begin{equation*}
    \Lambda(0,T) = [y(T) -r_0\Gamma]. 
\end{equation*}
This gives an ordinary differential equation (ODE) for $a(t)$
\begin{equation}
\begin{aligned}
\frac{\partial}{\partial t}a(t) e^{2\kappa(t)} + 2a(t) \cdot b(t) \cdot e^{2\kappa(t)} - e^{2\kappa(t)}\sigma^2 (t)\\
= \frac{\partial}{\partial t} \Bigg [ e^{\kappa(t)} \Bigg [ e^{\kappa(t)} \frac{\partial}{\partial t}(y(t) - r_0\Gamma(0,t)) \Bigg ]\Bigg ],
\label{21}
\end{aligned}
\end{equation}
where $y(T)$ is the simulated yield rate at tenor point $T$.
We can solve \eqref{21} numerically with the given initial conditions and yield rates for $0\leq T_m \leq T$. For all $T_m \in [0, T]$, we know $b,\sigma$ and $\Gamma(0,T)$ from the given initial conditions and (\ref{18}) respectively.
If we assume $a(t)$ to be piecewise constant with values $a(i)$ in $((i+1).\Delta T,i.\Delta T)$, then we can calculate $a(i)$ iteratively for $0 \leq T_m \leq T$. This leads to a triangular system of linear equations for the vector $a(i)$ with non-zero diagonal elements.
\begin{equation}
    Ea = F,
    \label{22}
\end{equation}
where $E$ maps the parameter $a(t)$ of the Hull-White model based on the simulated yield curves obtained using the market data. 
The authors of \cite{Engl07} noticed that a small change in the market data used to obtain the yield curves leads to large disturbances in the model parameter $a(t)$. This makes the problem of solving a system of linear equations ill-posed. We can define a well-posed problem with the following three different properties.
\theoremstyle{definition}
\begin{definition}
\textbf{A well-posed problem} \cite{Hadamard1902} \textit{A given problem is said to be well-posed if it holds the following three properties.
\begin{enumerate}
    \item For all suitable data, a solution exist,
    \item has an unique solution, and
    \item the solution changes continuously with data.
\end{enumerate}}
\end{definition}
However, in our case, a small change in the market data may lead to the large perturbations in the model parameters, and which violates the third property. Hence, the use of naive approaches may result in some numerical errors.
To regularize the ill-posed problem, we implement Tikhonov regularization.
\begin{equation}
        a^\delta_\mu = \mathrm{argmin}\|Ea - F^\delta \|^2 + \mu\|a\|^2
        \label{23}
\end{equation}
where $a^\delta_\mu$ is an estimate for $a$, $\mu$ is the regularization parameter, $\delta = \| F - F^\delta\|$ is the noise level, and $\mu\|a\|^2$ is a penalty term. We solve the optimization problem (\ref{23}) to obtain the parameter $a(t)$.
In this work, we use the commercially available software called UnRisk PRICING ENGINE for the parameter calibrations \cite{MathConsult09}. The UnRisk engine implements a classical Tikhonov regularization approach to regularize the ill-posed problems. By providing the simulated yield curve, the UnRisk pricing function returns the calibrated parameter $a(t)$ for that yield curve.
Based on 10,000 different simulated yield curves, we obtain $s=$10,000 different parameter vectors $a(t)$. In the matrix form, we write
\begin{equation}
    \mathfrak{A} =
\begin{bmatrix}
a_{11} & \cdots & a_{1m}\\ 
\vdots & \vdots & \vdots\\
a_{s1} & \cdots & a_{sm}\\ 
\end{bmatrix}
\label{24}
\end{equation}
where $m$ is the number of tenor points. All parameters $a_{i,j}$ are assumed to be piecewise constant changing their values only on tenor points, i.e., $m$ tenor points mean $m$ values for a single parameter vector.
\clearpage
\section{Numerical Methods}
\label{NumericalMethod}
\begin{figure}[H]
\begin{center}
\usetikzlibrary{fit}
\usetikzlibrary{shapes,arrows,positioning}
\tikzstyle{decision} = [diamond, draw, minimum height=3em, text centered]
\tikzstyle{line}=[draw]
\tikzstyle{block} = [rectangle, draw,
    text width=8cm, text centered, rounded corners, minimum height=3em]
\tikzstyle{line} = [draw, -latex']
\tikzstyle{container} = [draw, rectangle, dashed, inner sep=0.3cm]
\begin{tikzpicture}[node distance = 2.2cm, auto]
\node[block](PDE){Hull-White model
\begin{equation*}
    \frac{\partial V}{\partial t} + (a(t) - br)\frac{\partial V}{\partial r} + \frac{1}{2} \sigma^2 \frac{\partial^2 V}{\partial r^2} - rV = 0
\end{equation*}
};
\node[block, below of=PDE](FDM){Finite difference method
\begin{equation*}
    A(\rho_s(t))V^{n+1} = B(\rho_s(t))V^{n}, \hspace{1cm} V(0) = V_0
\end{equation*}
};
\node[block, below of=FDM](CG){Selection of training parameters: \\Classical greedy sampling algorithm\\
$\rho_1,...,\rho_l$};
%\node[block, right of=dec1, xshift=5cm, text width=3cm](AG){Adaptive greedy sampling\\ algorithm};
\node[block, below of= CG](POD){POD:
Method of snapshots\\
$\hat{V} = [V(\rho_1),V(\rho_2),...,V(\rho_l)]$
};
\node[block, below of=POD](SVD){Singular value decomposition
\begin{equation*}
   \hat{V} = \sum_{i=1}^k \Sigma_i\phi_i\psi_i^T.
\end{equation*}
};
\node[block, below of=SVD, yshift=-0.5cm](ROM){Reduced order model
\begin{equation*}
    A_d(\rho_s)V^{n+1}_d = B_d(\rho_s)V^n_d
\end{equation*}
};
%\node [decision, text width=2.5cm, below of=CG, yshift=-1.5cm](dec1){
%ROM Quality
%\begin{equation*}
   % \|e\|_2^2 = \frac{\|{V - \Bar{V}}\|_2^2}{\| V\|_2^2}
%\end{equation*}
%};
\node [container,fit= (POD) (SVD) (ROM),
] (container) {};
\node at (container.north) [above right,node distance=0 and 0,font=\fontsize{12}{0}\selectfont] {\textbf{Model Order Reduction}};
\node [decision, text width=2.5cm, below of=ROM, yshift=-1.0cm](dec1){
ROM Quality
};
\node[block, right of=dec1, xshift=5cm, text width=5cm](AG){Adaptive greedy sampling\\ algorithm};
\node[block, below of = AG, text width=5cm](MOR){\textbf{Model Order Reduction}};
\node[block, below of = dec1, text width=5cm, yshift=-1.0cm](S1){Stop};
\node[block, below of = MOR, text width=5cm](S2){Stop};
%\node[above] at (container.north) {Caption};
\path [line] (PDE) -- (FDM);
\path [line] (FDM) -- (CG);
%\path [line] (CG) -- (TP);
\path [line] (CG) -- (POD);
%\path[line] (dec1) -- node[anchor=east] {satisfactory} (POD);
\path [line] (POD) -- (SVD);
\path [line] (SVD) -- (ROM);
\path [line] (ROM) -- (dec1);
\path [line] (dec1) -- node[anchor=east] {satisfactory} (S1);
\path [line] (dec1) -- node[anchor=south] {unsatisfactory} (AG);
\path [line] (AG) -- (MOR);
\path [line] (MOR) -- (S2);
%\path [line] (dec1) -- node[anchor=south] {unsatisfactory} (AG);
%\draw [->] (AG) |- (POD);
\end{tikzpicture}
\end{center}
\caption{Model hierarchy to obtain a reduced order model for the Hull-White model.}
\label{fig:2}
\end{figure}
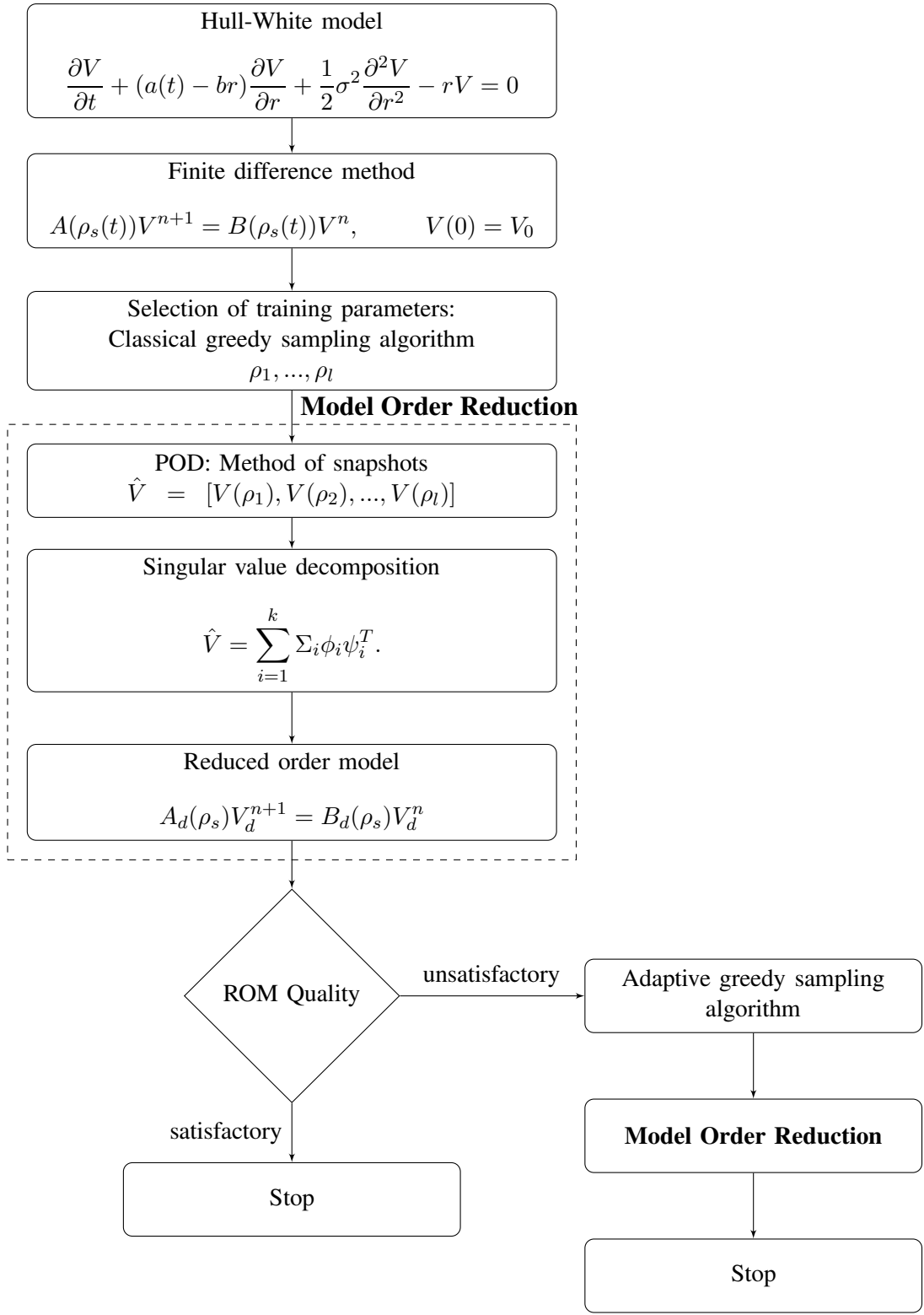

Figure \ref{fig:2} shows the model hierarchy to obtain a reduced-order model for the Hull-White model.
We discretize the Hull-White PDE using a finite difference method as presented in the subsection \ref{FDMHW}. The discretization of the PDE creates a parameter-dependent high dimensional model. We need to solve the high dimensional model for at least 10,000 parameters calibrated in the previous subsection \ref{ParaCalb}. Solving the high dimensional model for such a large parameter space is computationally costly. Thus, we incorporate the parametric model order reduction approach based on the proper orthogonal decomposition. The POD approach relies on the method of snapshots. The snapshots are nothing but the solutions of the high dimensional model at some parameter values.\\
The idea is to solve the high dimensional model for only a certain number of training parameters to obtain a reduced-order basis. This reduced-order basis is then used to construct a reduced-order model. Subsection \ref{MORPOD} illustrates the proper orthogonal decomposition approach, along with the construction of snapshots, and the Algorithm \ref{Algo1} presents the methodology to obtain the reduced-order basis. Finally, we can solve the reduced-order model cheaply for the large parameter space. The selection of the training parameters is of utmost importance to obtain the optimal reduced-order model. In subsection \ref{CG}, we introduce a greedy sampling technique for the selection of training parameters. The greedy sampling technique selects the parameters at which the error between the reduced-order model and the high dimensional model is maximal. However, the greedy sampling approach exhibits some drawbacks (see subsection \ref{CGDrawbacks}). To avoid these drawbacks, in the subsection \ref{AG}, we implement an adaptive greedy approach. These methods are interlinked with each other and necessary to obtain the most efficient reduced-order model.
\subsection{Finite Difference Method}
\label{FDM}
The PDEs obtained for the Hull-White model is a convection-diffusion-reaction type PDE \cite{Anderson13}. Consider a Hull-White PDE given by (\ref{20})
\begin{equation}
    \frac{\partial V}{\partial t} + \underbrace{(a(t) - br_t) \frac{\partial V}{\partial r}}_\text{Convection} + \underbrace{\frac{1}{2} \sigma^2 \frac{\partial^2 V}{\partial r^2}}_\text{diffusion} - \underbrace{rV}_\text{reaction} = 0.
    \label{25}
\end{equation}
In this work, we apply a finite difference method to solve the Hull-White PDE. The convection term in the above equation may lead to numerically unstable results. Thus, we implement the so-called upwind scheme \cite{Heinrich77} to obtain a stable solution. We incorporate the semi-implicit scheme called the Crank-Nicolson method \cite{Sun06} for the time discretization.
\subsubsection{Spatial Discretization}
\label{S-mesh}
Consider the following one-dimensional linear advection equation for a field $\zeta(x,t)$,
\begin{equation}
    \frac{\partial \zeta}{\partial t} + U \frac{\partial \zeta}{\partial x} = 0,
    \label{26}
\end{equation}
describing a wave propagation along the $x-$axis with a velocity $U$.
We define a discretization of the computational domain in $sd$ spatial dimensions as
\begin{equation*}
        [u_k,v_k]^{sd} \times [0, T] = [u_1,v_1] \times \cdots \times [u_{sd},v_{sd}] \times [0,T] = \bigg ( \prod_{k=1}^{sd} [u_k,v_k] \bigg)\times [0,T],
\end{equation*}
where $u$ and $v$ are the cut off limits of the spatial domain. $T$ denotes the final time of the computation. The corresponding indices are $i_k \in \{ 1, \dots, M_k\}$ for the spatial discretization and $n \in \{1,\dots ,N\}$ for the time discretization.
The first order upwind scheme of order $\mathcal{O}(\Delta x)$ is given by
\begin{subequations}
\begin{align}
    \frac{\zeta_i^{n+1} - \zeta_i^n}{\Delta t} + U \frac{\zeta_i^n - \zeta_{i-1}^n}{\Delta x} &= 0 \hspace{2cm} \, \, \text{for } U >0 \\
    \frac{\zeta_i^{n+1} - \zeta_i^n}{\Delta t} + U \frac{\zeta_{i+1}^n - \zeta_{i}^n}{\Delta x} &= 0 \hspace{2cm} \, \, \text{for } U <0
\end{align}
\label{27}
\end{subequations}
Let us introduce,
\begin{equation*}
    U^+ = \mathrm{max}(U,0), \hspace{2cm} U^- = \mathrm{min}(U,0),
\end{equation*}
and
\begin{equation*}
    \zeta_x^- = \frac{\zeta_i^n - \zeta_{i-1}^n}{\Delta x}, \hspace{2cm} \zeta_x^+ = \frac{\zeta_{i+1}^n - \zeta_{i}^n}{\Delta x}
\end{equation*}
Combining (\ref{27}a) and (\ref{27}b) in compact form, we obtain
\begin{equation}
    \zeta_i^{n+1} = \zeta_i^n - \Delta t[U^+ \zeta_x^- + U^- \zeta_x^+].
    \label{28}
\end{equation}
We have implemented the above defined upwind scheme for the convection term. The diffusion term is discretized using the second order central difference scheme of order $\mathcal{O}(\Delta x)^2$ given by
\begin{equation}
    \frac{\partial^2 \zeta}{\partial x^2} = \frac{\zeta_{i+1}^n - 2\zeta_i^n + \zeta_{i-1}^n}{(\Delta x)^2}.
    \label{29}
\end{equation}
\subsubsection{Time Discretization}
\label{t-mesh}
Consider a time-dependent PDE for a quantity $\zeta$
\begin{equation}
    \frac{\partial \zeta}{\partial t} + L \zeta = 0,
    \label{30}
\end{equation}
where $L$ is the differential operator containing all spatial derivatives. Using the Taylor series expansion, we write
\begin{equation*}
    \zeta(t+\Delta t) = \zeta(t) + \Delta \frac{\partial \zeta}{\partial t} + \frac{\Delta t^2}{2} \frac{\partial^2 \zeta}{\partial t^2}.
\end{equation*} 
Neglecting terms of order higher than one, we obtain
\begin{equation*}
    \frac{\partial \zeta}{\partial t} = \frac{\zeta (t+\Delta t) - \zeta(t)}{\Delta t} + \mathcal{O}(\Delta t).
\end{equation*}
From (\ref{30}), we get
\begin{equation}
    \zeta(t+\Delta t) = \zeta(t) - \Delta t (L(t) \zeta(t)).
    \label{31}
\end{equation}
Let us introduce a new parameter $\Theta$ such that
\begin{equation}
     \frac{\zeta(t+\Delta t) - \zeta(t)}{\Delta t} = (1-\Theta)(L(t) \zeta(t)) + \Theta(L(t+ \Delta t) \zeta(t + \Delta t)).
     \label{32}
\end{equation}
We can construct different time discretization schemes for different values of $\Theta$. Setting $\Theta = 0$, we obtain a fully explicit scheme known as the forward difference method, while considering $\Theta = 1$, we get a fully implicit scheme known as the backward difference method. Here we set $\Theta = 1/2$ and obtain a semi-implicit scheme known as the Crank-Nicolson method \cite{Anderson13}
\begin{equation}
    \bigg ( 1 - \frac{1}{2}\Delta t L(t+\Delta t) \bigg)\zeta(t+\Delta t) = \bigg ( 1 + \frac{1}{2}\Delta t L(t) \bigg)\zeta(t).
    \label{33}
\end{equation}
\subsubsection{Finite Difference Method for a Hull-White Model}
\label{FDMHW}
The computational domain for a spatial dimension the rate $r$ is $[u,v]$. According to \cite{Binder013}, the cut off values $u$ and $v$ are given as
\begin{equation}
    u = r_{sp} + 7\sigma \sqrt{T}  \text{      and      } v = r_{sp} - 7\sigma \sqrt{T},
    \label{34}
\end{equation}
where $r_{sp}$ is a yield at the maturity $T$ also known as a spot rate.
We divide the spatial domain into $M$ equidistant grid points which generate a set of points $\{r_1,r_2,\dots,r_M \}$. The time interval $[0,T]$ is divided into $N-1$ time points ($N$ points in time that are measured in days starting from $t=0$).\\
Equation (\ref{25}) can then be represented as,
\begin{equation}
    \frac{\partial V}{\partial t} + L(t)V(t) = 0.
    \label{35}
\end{equation}
We specify the spatial discretization operator $L(n)$, where the index $n$ denotes the time-point.
From (\ref{27}) and (\ref{29}), we get,
\begin{equation}
    \begin{aligned}
    \text{for}\,\, (a(n) - br_i) > 0\\
    L(n)V_i^n &:=
    &\frac{1}{2}\sigma^2 \frac{V_{i+1}^n - 2V_i^n + V_{i-1}^n}{(\Delta x)^2} + (a(n) - br_i)\frac{V_i^n - V_{i-1}^n}{\Delta x} - r_iV_i^n,\\
    \text{for}\,\, (a(n) - br_i) < 0\\
    L(n)V_i^n &:=
    &\frac{1}{2}\sigma^2 \frac{V_{i+1}^n - 2V_i^n + V_{i-1}^n}{(\Delta x)^2} + (a(n) - br_i)\frac{V_{i+1}^n - V_i^n}{\Delta x} - r_iV_i^n.\\
    \end{aligned}
    \label{36}
\end{equation}
From (\ref{33}), we obtain
\begin{equation}
    \frac{V(t+\Delta t)-V(t)}{\Delta t} = (1-\Theta)(L(t) V(t)) + \Theta(L(t+ \Delta t) V(t + \Delta t))
    \label{37}
\end{equation}
For $\Theta = 1/2$, we then have
 \begin{equation}
    \underbrace{\bigg ( 1 - \frac{1}{2}\Delta t L(t+\Delta t) \bigg)}_{A(\rho_s(t)) \in \mathbb{R}^{M\times M}} V(t+\Delta t) = \underbrace{\bigg ( 1 + \frac{1}{2}\Delta t L(t) \bigg)}_{B(\rho_s(t)) \in \mathbb{R}^{M\times M}} V(t),
    \label{38}
 \end{equation}
where the matrices $A(\rho_s(t))$, and $B(\rho_s(t))$ depend on parameters $a(t)$, $b$ and $\sigma$. We denote $\rho_s = \{a(t),b,\sigma\}$ as the $s$th group of these parameters.
Here
\begin{equation*}
    A(\rho_s(t)) = I - \frac{\sigma^2\Delta t}{2(\Delta x)^2}J - \frac{\Delta t}{2\Delta x}(H^+G^- + H^-G^+) + R_o,
\end{equation*}
and
\begin{equation*}
    B(\rho_s(t)) = I + \frac{\sigma^2\Delta t}{2(\Delta x)^2}J + \frac{\Delta t}{2\Delta x}(H^+G^- + H^-G^+) - R_o,
\end{equation*}
where
\begin{equation*}
    J =
\begin{bmatrix}
-2 & 1 & 0 & \cdots & 0\\ 
1 & -2 & 1 & \ddots & \vdots\\ 
0 & 1 & \ddots & \ddots & 0\\ 
\vdots &\ddots & \ddots & \ddots & 1\\  
0 & \cdots & 0 & 1 & -2\\
\end{bmatrix}
\end{equation*}
\begin{equation*}
    G^- =
\begin{bmatrix}
1 & 0 & 0 & \cdots & 0\\ 
-1 & 1 & 0 & \ddots & \vdots\\ 
0 & -1 & \ddots & \ddots & 0\\ 
\vdots &\ddots & \ddots & \ddots & 0\\  
0 & \cdots & 0 & -1 & 1\\
\end{bmatrix}\\
  \,\,\,\,\,  G^+ =
\begin{bmatrix}
1 & -1 & 0 & \cdots & 0\\ 
0 & 1 & -1 & \ddots & \vdots\\ 
0 & 0 & \ddots & \ddots & 0\\ 
\vdots &\ddots & \ddots & \ddots & -1\\  
0 & \cdots & 0 & 0 & 1\\
\end{bmatrix}
\end{equation*}
\begin{equation*}
    H^+ =
\begin{bmatrix}
\mathrm{max}(a(n) - br(1)) & 0 & 0 & \cdots & 0\\ 
0 & \mathrm{max}(a(n) - br(2)) & 0 & \ddots & \vdots\\ 
0 & 0 & \ddots & \ddots & 0\\ 
\vdots &\ddots & \ddots & \ddots & 0\\  
0 & \cdots & 0 & 0 & \mathrm{max}(a(n) - br(M))\\
\end{bmatrix}
\end{equation*}
\begin{equation*}
    H^- =
\begin{bmatrix}
\mathrm{min}(a(n) - br(1)) & 0 & 0 & \cdots & 0\\ 
0 & \mathrm{min}(a(n) - br(2)) & 0 & \ddots & \vdots\\ 
0 & 0 & \ddots & \ddots & 0\\ 
\vdots &\ddots & \ddots & \ddots & 0\\  
0 & \cdots & 0 & 0 & \mathrm{min}(a(n) - br(M))\\
\end{bmatrix}
\end{equation*}
\begin{equation*}
    R_o =
\begin{bmatrix}
r(1) & 0 & 0 & \cdots & 0\\ 
0 & r(2) & 0 & \ddots & \vdots\\ 
0 & 0 & \ddots & \ddots & 0\\ 
\vdots &\ddots & \ddots & \ddots & 0\\  
0 & \cdots & 0 & 0 & r(M)\\
\end{bmatrix}
\end{equation*}
The above discretization of the PDE generates a parametric high dimensional model of the following form (\ref{39}).
\begin{equation}
    A(\rho_s(t))V^{n+1} = B(\rho_s(t))V^{n}, \hspace{2cm} V(0) = V_0,
    \label{39}
\end{equation}
where the matrices $A(\rho) \in \mathbb{R}^{M\times M}$, and $B(\rho) \in \mathbb{R}^{M\times M}$ are parameter dependent matrices. $M$ is the total number of spatial discretization points. $t$ is the time variable. $t= [0,T]$ where $T$ is the final term date.
For the simplicity of notations, we denote $\rho_s=\{(a_{s1},\dots,a_{sm}),b,\sigma\}$ as the $s$th group of model parameters where $s = 1,\dots,10000$, and $m$ is the total number of tenor points. We need to solve this system at each time step $n$ with an appropriate boundary condition and a known initial value of the underlying instrument.
We need to solve the system (\ref{39}) for at least 10,000 parameter groups $\rho$ generating a parameter space $P$ of $10000 \times m$.
\begin{table}[htb]
\caption{List of symbols used in subsection \ref{FDMHW}}
\label{tab:1}
\setlength{\tabcolsep}{1cm}
\begin{center}
\begin{tabular}{ll}
\hline\noalign{\smallskip}
$M$ & Spatial discretization points. \\
$N$ & Temporal discretization points. \\
$A$, $B$ & System matrices. \\
$T$ & Maturity or the final term date. \\
$m$ & Number of tenor points. \\
$\rho$ & Group of model parameters $\{a(t),b,\sigma\}$. \\
$P$ & Parameter space.\\
\noalign{\smallskip}\hline
\end{tabular}
\end{center}
\end{table}
\clearpage
\subsection{Parametric Model Order Reduction}
\label{MORPOD}
We employ the projection based model order reduction (MOR) technique to reduce the high dimensional model (\ref{39}). The reduced-order model is obtained using the Galerkin projection onto the low dimensional subspace, $Q \in \{\phi_i\}_{i=1}^{d}$. We approximate the high dimensional state space $V^n$ by a Galerkin ansatz as
\begin{equation}
    \Bar{V}^n = QV^n_d,
    \label{40}
\end{equation}
where $Q \in \mathbb{R}^{M\times d}$ is the reduced-order basis with $d \ll M$, $V_d$ is a vector of reduced coordinates, and $\Bar{V} \in \mathbb{R}^{M}$ is the solution obtained using the reduced order model.
Substituting (\ref{40}) into the system of equations (\ref{39}) gives the residual of the reduced state as
\begin{equation}
    R^n(V_d,\rho_s) = A(\rho_s)QV^{n+1}_d - B(\rho_s)QV^n_d.
    \label{41}
\end{equation}
In the case of the Galerkin projection, the residual $R(V_d,\rho_s)$ is orthogonal to the ROB $Q$
\begin{equation}
    Q^TR^n(V^n_d,\rho_s) = 0.
    \label{42}
\end{equation}
Multiplying (\ref{41}) by $Q^T$, we get
\begin{equation}
\begin{aligned}
    Q^TA(\rho_s)QV^{n+1}_d &= Q^TB(\rho_s)QV^n_d,\\
    A_d(\rho_s)V^{n+1}_d &= B_d(\rho_s)V^n_d,
    \label{43}
\end{aligned}
\end{equation}
where the matrices $A_d(\rho_s) \in \mathbb{R}^{d\times d}$ and $B_d(\rho_s) \in \mathbb{R}^{d \times d}$ are the parameter dependent reduced matrices. In short, instead of solving a linear system of equations of order $M$, the MOR approach solves a linear system of order $d$ where $d \ll M$. We obtain the Galerkin projection matrix $Q$ (\ref{43}) based on a proper orthogonal decomposition (POD) method. POD generates an optimal order orthonormal basis $Q$ in the least square sense which serves as the reduced-order basis for a given set of computational data. We aim to obtain the subspace $Q$ independent of the parameter space $P$. In this work, we obtain the reduced-order basis by the method of snapshots.
The snapshots are nothing but the state solutions obtained by solving the high dimensional models for selected parameter groups. We assume that we have a training set of parameter groups $\rho_{1}$ $\cdots$ $\rho_{l}$ $\in$ $[\rho_{1}$ $\rho_{s}]$. We compute the solutions of the high dimensional models for the training set and combine them to form a snapshot matrix $\hat{V} = [V(\rho_1),V(\rho_2),...,V(\rho_l)]$. Now, to obtain the desired reduced-order basis, the POD method solves
\begin{equation}
    \mathrm{POD}(\hat{V}) := \underset{Q}{\mathrm{argmin}} \frac{1}{l} \sum_{i=1}^{l} \|V_i - QQ^TV_i\|^2,
    \label{44}
\end{equation}
for all matrices $Q \in \mathbb{R}^{M \times d}$ that satisfy $QQ^T = I$. We can obtain the reduced-order basis $Q$ fulfilling the condition (\ref{44}) by computing the SVD of the matrix $\hat{V}$. 
We perform a truncated SVD \cite{Golub70} of the matrix $\hat{V}$ to obtain the reduced-order basis $Q$
\begin{equation}
\begin{aligned}
   \hat{V} &= \sum_{i=1}^k \Sigma_i\phi_i\psi_i^T,\\
   \hat{V} &= \Phi \Sigma \Psi^T,
    \end{aligned}
    \label{45}
\end{equation}
where $\phi_i$ and $\psi_i$ are the left and right singular vectors of the matrix $\hat{V}$ respectively, and $\Sigma_i$ are the singular values.
\begin{equation*}
\hat{V}= 
\begin{bmatrix}
\phi_1 & \cdots & \phi_k\\
\end{bmatrix}_{M\times k}
\begin{bmatrix}
\Sigma_1 & 0 & \cdots\\ 
0 & \ddots & \vdots\\ 
\vdots & \cdots & \Sigma_k\\
\end{bmatrix}_{k \times k}
\begin{bmatrix}
\psi_{1} & \cdots & \psi_{k}\\
\end{bmatrix}_{k \times k}
\label{71}
\end{equation*}
The truncated (economy-size) SVD computes only the first $k$ columns of the matrix $\Phi$. 
The optimal projection subspace $Q$ consists of $d$ left singular vectors $\phi_i$ known as POD modes. Here $d$ is the dimension of the reduced order model.\\
The Algorithm \ref{Algo1} shows the steps to construct a reduced-order basis using the proper orthogonal decomposition approach.
\begin{table}[htb]
\caption{List of symbols used in the Algorithm \ref{Algo1}}
\label{tab:2}
\setlength{\tabcolsep}{1cm}
\begin{center}
\begin{tabular}{ll}
\hline\noalign{\smallskip}
$a(t)$ & Deterministic drift.\\
$b$ & Mean reversion speed. \\
$\sigma$ & Volatility. \\
$\rho$ & Group of model parameters $\{a(t),b,\sigma\}$. \\
$V(\rho_i)$ & Solution obtained using a high dimensional model for $\rho_i$.\\
$\hat{V}$ & Snapshot matrix. \\
$\Phi$ & Matrix of left singular vectors. \\
$\Psi$ & Matrix of right singular vectors. \\
$\Sigma$ & Matrix of singular values. \\
$\Xi_j$ & Relative energy of the $j$th POD mode.\\
$Q$ & Reduced-order basis. \\
HDM & High dimensional model. \\
ROM & Reduced-order model.\\
\noalign{\smallskip}\hline
\end{tabular}
\end{center}
\end{table}
\begin{algorithm}
\caption{Reduced-order basis using a proper orthogonal decomposition}
\begin{algorithmic}[1]
\INPUT{Parameter $a(t)$, $b$, $\sigma$, Energy level $EL$, $l$}
\OUTPUT{$Q$}
\STATE{Choose $\rho_1,\cdots,\rho_l$}
\FOR{$i = 1$ to $l$}
\STATE{Solve the HDM for the parameter group $\rho_i$ : $V(\rho_i)$}
\ENDFOR
\STATE{Construct a snapshot matrix $\hat{V}$ using $\{V(\rho_i)\}_{i =1}^{l}$}
\STATE{$\hat{V} = [V(\rho_1),...,V(\rho_l)]$}
\STATE{Compute the leading singular values and associated singular vectors of $\hat{V}$ using the truncated SVD: $\hat{V}= \Phi \Sigma \Psi^T$}
\STATE{$\Xi=\mathrm{diag}(\Sigma)/\mathrm{sum}(\mathrm{diag}(\Sigma))$}
\FOR{$j = 1$ to $\mathrm{length}(\Sigma)$}
\STATE{$\Bar{\Xi} = \mathrm{sum}(\Xi(1:j))\times 100$}
\IF{$\Bar{\Xi} >$ $EL$}
\STATE{$d = j$}
\ENDIF
\ENDFOR
\STATE{$Q = [\phi_1 \cdots \phi_{d}]$}
\end{algorithmic}
\label{Algo1}
\end{algorithm}
We have to choose the dimension $d$ of the subspace $Q$ such that we get a good approximation of the snapshot matrix. According to \cite{Rathinam03}, large singular values correspond to the main characteristics of the system, while small singular values give only small perturbations of the overall dynamics. The relative importance of the $i$th POD mode of the matrix $\hat{V}$ is determined by the \emph{relative energy} $\Xi_i$ of that mode
\begin{equation}
    \Xi_i = \frac{\Sigma_i}{\sum_{i=1}^k \Sigma_i} 
    \label{46}
\end{equation}
If the sum of the energies of the generated modes is unity, then these modes can be used to reconstruct a snapshot matrix completely \cite{Williams13}. Generally, the number of modes required to generate the complete data set is significantly less that the total number of POD modes \cite{Pinnau08}. Thus, a matrix $\hat{V}$ can be accurately approximated by using POD modes whose corresponding energies sum to almost all of the total energy. 
Thus, we choose only $d$ out of $k$ POD modes to construct $Q = [\phi_1 \cdots \phi_{d}]$ which is a parameter independent projection subspace based on (\ref{46}).
It is evident that the quality of the reduced-order model mainly depends on the selection of parameter groups $\rho_1,...,\rho_l$ use to compute snapshots. Thus, it necessitates defining an efficient sampling technique for the high dimensional parameter space. We can consider the standard sampling techniques like uniform sampling or random sampling \cite{James013}. However, these techniques may neglect a vital region within the parameter space. Alternatively, the greedy sampling method is proven to be an efficient method for sampling a high dimensional parameter space in the framework of model order reduction \cite{Grepl05,Paul14,Amsallem015}.
\subsection{Greedy Sampling Method}
\label{CG}
The greedy sampling technique selects the parameter groups at which the error between the reduced-order model and the high dimensional model is maximal. Further, we compute the snapshots using these parameter groups so that we can obtain the best suitable reduced-order basis $Q$.
\begin{equation}
\begin{aligned}
    \|e\| = \frac{\|{V - \Bar{V}}\|}{\| V\|}.\\
    \rho_I = \underset{\rho \in P}{\mathrm{argmax}} \;\|e\|,
\end{aligned}
\label{47}
\end{equation}
where $\|e\|$ is a relative error between the reduced-order model and the high dimensional model. Thus, the greedy sampling algorithm, at each greedy iteration $i = 1,...,I_{max}$, selects the optimal parameter group $\rho_I$, which maximizes the relative error $\|e\|$.
However, the computation of relative error $ \|{e}\|$ is costly as it entails the solution for the high dimensional model. Thus, usually, the relative error is replaced by error bounds or the residual error $\|R\|$. However, in some cases, it is not possible to define the error bounds or the error bounds do not exist. In such cases, the norm of the residual is a good alternative \cite{Thanh08,Veroy05,Paul14}. For the simplicity of notation, we consider $\varepsilon$ as the error estimator, i.e., in our case the norm of the residual. The greedy sampling algorithm runs for $I_{max}$ iterations. At each iteration $i = 1,...,I_{max}$, we choose the parameter group as the maximizer
\begin{equation}
    \rho_I = \underset{\rho \in P}{\mathrm{argmax}} \;\varepsilon(\rho).
    \label{48}
\end{equation}
The Algorithm \ref{Algo2} describes the classical greedy sampling approach. It initiates by selecting any parameter group $\rho_1$ from the parameter set $P$ and computes a reduced-order basis $Q_1$, as described in subsection \ref{MORPOD}. It is necessary to note that the choice of a first parameter group to obtain $Q_1$ does not affect the final result. Nonetheless, for the simplicity of computations, we select the first parameter group $\rho_1$ from the parameter space $P$.
Furthermore, the algorithm chooses the pre-defined parameter set $\hat{P}$ randomly of cardinality $C$ as a subset of $P$. At each point within the parameter set $\hat{P}$, the algorithm determines a reduced-order model using the reduced-order basis $Q_1$ and computes error estimator values, $\varepsilon(\rho_j)_{j=1}^C$. The parameter group in the pre-defined parameter set $\hat{P}$ at which the error estimator is maximal is then selected as the optimal parameter group $\rho_I$. Furthermore, the algorithm solves the high dimensional model for the optimal parameter group and updates the snapshot matrix $\hat{V}$.
Finally, a new reduced-order basis is obtained by computing a truncated singular value decomposition of the updated snapshot matrix, as described in the Algorithm \ref{Algo1}. These steps are then repeated for $I_{max}$ iterations or until the maximum value of the error estimator is higher than the specified tolerance $\varepsilon_{tol}$.
\begin{table}[htb]
\caption{List of symbols used in the Algorithm \ref{Algo2}}
\label{tab:3}
\setlength{\tabcolsep}{1cm}
\begin{center}
\begin{tabular}{ll}
\hline\noalign{\smallskip}
$I_{max}$ & Maximum number of greedy iterations.\\
$C$ & Maximum parameter groups selected to obtain a reduced-order basis. \\
$P$ & Parameter space. \\
$\rho$ & Group of model parameters $\{a(t),b,\sigma\}$. \\
$Q$ & Reduced-order basis (ROB). \\
$\varepsilon$ & Error estimator. \\
$\varepsilon_{tol}$ & Tolerance for the error estimator, greedy iteration terminates if $\varepsilon < \varepsilon_{tol}$. \\
$V(\rho_i)$ & Solution obtained by solving a high dimensional model for $\rho_i$.\\
$\hat{V}$ & Snapshot matrix. \\
\noalign{\smallskip}\hline
\end{tabular}
\end{center}
\end{table}
\begin{algorithm}[H]
\caption{The classical greedy sampling algorithm}
\begin{algorithmic}[1]
\INPUT{Maximum number of iterations $I_{max}$, maximum parameter groups $C$, Parameter space $P$, $\varepsilon_{tol}$}
\OUTPUT{$Q$}
\STATE{Choose first parameter group $\rho_1 = [(a_{11},...,a_{1m}),b,\sigma]$ from $P$}
\STATE{Solve the HDM for a parameter group $\rho_1$\ and store the results in $V_1$}
\STATE{Compute a truncated SVD of the matrix $V_1$ and construct $Q_1$}
\STATE{Randomly select a set of parameter groups $\hat{P} = \{ \rho_1, \rho_2,..., \rho_C\} \subset P$}
\FOR{i = 2 to $I_{max}$}
    \FOR{j = 1 to $C$}
        \STATE{Solve a ROM for the parameter group $\rho_j$ with the ROB $Q_{i-1}$}
        \STATE{Compute the error estimator $\bm{\varepsilon}(\rho_j)$}
    \ENDFOR
    \STATE{Find $\rho_I = \underset{\rho \in \hat{P}}{\mathrm{argmax}} \;\bm{\varepsilon}(\rho)$}
    \IF{$\varepsilon(\rho_I) \leq \bm{\varepsilon}_{tol}$}
        \STATE{$Q = Q_{i-1}$}
        \STATE{\textbf{break}}
    \ENDIF
    \STATE{Solve the HDM for the parameter group $\rho_I$ and store the result in $V_i$}
    \STATE{Construct a snapshot matrix $\hat{V}$ by concatenating the solutions $V_s$ for $s = 1,...,i$}
    \STATE{Compute an SVD of the matrix $\hat{V}$ and construct $Q_i$}
\ENDFOR
\end{algorithmic}
\label{Algo2}
\end{algorithm}
\subsubsection{Drawbacks}
\label{CGDrawbacks}
The greedy sampling method computes an inexpensive \textit{a posteriori} error estimator for the reduced-order model.
However, it is not feasible to calculate the error estimator values for the entire parameter space $P$.
An error estimator is based on the norm of the residual which scales with the dimension of the high dimensional model, $M$. With an increase in dimension, it is not computationally reasonable to calculate the residual for 10,000 parameter groups, i.e., to solve the reduced-order model for the entire parameter space. Hence, the classical greedy sampling technique chooses the pre-defined parameter set $\hat{P}$ randomly as a subset of $P$. Random sampling is designed to represent the whole parameter space $P$, but there is no guarantee that $\hat{P}$ will reflect the complete space $P$ since the random selection of a parameter set may neglect the parameter groups corresponding to the most significant error. These observations motivate to design a new criterion for the selection of the subset $\hat{P}$.\\
Another drawback of the classical greedy sampling technique is that we have to specify the maximum error estimator tolerance $\varepsilon_{tol}$. The error estimator usually depends on some error bound, which is not tight or it may not exist. To overcome this drawback, we establish a strategy by constructing a model for an exact error as a function of the error estimator based on the idea presented in \cite{Paul14}. Furthermore, we use this exact error model to observe the convergence of the greedy sampling algorithm instead of the error estimator.
\subsection{Adaptive Greedy Sampling Method}
\label{AG}
To avoid the drawbacks associated with the classical greedy sampling technique, we have derived an adaptive greedy sampling approach which selects the parameter groups adaptively at each iteration of the greedy procedure, using an optimized search based on surrogate modeling. We construct a surrogate model of the error estimator $\Bar{\varepsilon}$ to approximate the error estimator $\varepsilon$ over the entire parameter space. Further, we use this surrogate model to locate the parameter groups $\hat{P_k} = \{\rho_1,...,\rho_{C_k}\}$ with $C_k < C$, where the values of the error estimator $\varepsilon$ are highest.
For each parameter group within the parameter set $\hat{P_k}$, we determine a reduced-order model and compute the error estimator values. The algorithm builds a new surrogate model based on these error estimator values, and the process repeats itself until the total number of parameter groups reaches $C$, resulting in the desired parameter set $\hat{P}$.
\subsubsection{Surrogate Modeling of the Error Estimator}
\label{SM}
At each greedy iteration, the algorithm construct a surrogate model of the error estimator to locate the parameters adaptively.
\begin{algorithm}[H]
\caption{Surrogate model using the principal component regression technique.}
\begin{algorithmic}[1]
\INPUT{The response vector $\bm{\hat{\varepsilon}} = [\varepsilon_1,...,\varepsilon_{C_k}]$, $\hat{P_k} = [\rho_1,...,\rho_{C_k}] \in \mathbb{R}^{C_k \times \Bar{m}}$}, principal components $p$
\OUTPUT{Vector of regression coefficients $\eta$}
\STATE{Standardize $\hat{P_k}$ and $\bm{\hat{\varepsilon}}$ with zero mean and variance equals to one}
\STATE{Compute the singular value decomposition of the matrix $\hat{P_k}$: \\
$\hat{P_k} = \hat{\Upphi}\hat{\Upsigma}\hat{\Uppsi}$
}
\STATE{Construct a new matrix $Z = \hat{P_k}\hat{\Uppsi}_p = [\hat{P_k}\hat{\uppsi}_1,...,\hat{P_k}\hat{\uppsi}_p]$ composed of principal components}
\STATE{Compute the least square regression using the principal components as independent variables:\\
$\hat{\Omega} = \underset{\Omega}{\mathrm{argmin}} \|\bm{\hat{\varepsilon}} - Z\Omega\|_2^2$}
\STATE{Compute the PCR estimate $\eta_{PCR}$ of the regression coefficients $\eta$: $\eta_{PCR} = \hat{\Uppsi}_p\hat{\Omega}$}
\end{algorithmic}
\label{Algo3}
\end{algorithm}
The detailed adaptive greedy sampling algorithm along with its description is presented in the subsection \ref{AGAlgo}. The first stage of the adaptive greedy sampling algorithm computes the error estimator over the randomly selected parameter set $\hat{P_0}$ of cardinality $C_0$. Furthermore, the algorithm uses these error estimator values $\{\varepsilon \}_{i=1}^{C_0}$ to build a surrogate model $\Bar{\varepsilon}_0$ and locates the $C_k$ parameter groups corresponding to the $C_k$ maximum values of the surrogate model. This process repeats itself for $k = 1,...,K$ iterations until the total number of parameter groups reaches $C$. Finally, the optimal parameter group $\rho_I$ is the one that maximizes the error estimator within the parameter set $\hat{P}$.
\begin{equation*}
        \hat{P} = \hat{P}_0 \cup \hat{P}_1 \cup \hat{P}_2 \cup \cdots \cup \hat{P}_{K}, \;\; \; \; k = 1,...,K
\end{equation*}
Thus, at each $k$th iteration, we construct a surrogate model $\Bar{\varepsilon}_k$ which approximates the error estimator over the entire parameter space $P$. There are different choices to build a surrogate model \cite{James013}. In this paper, we use the principal component regression (PCR) technique.
Suppose the vector $\bm{\hat{\varepsilon}} = (\varepsilon_1,...,\varepsilon_{C_k}) \in \mathbb{R}^{C_k \times 1}$ is the response vector having error estimator values at $k$th iteration. Since, we have considered parameters $b$ and $\sigma$ as constants, we build a surrogate model with the parameter $a(t)$ only. Let $\hat{P_k} = [\rho_1,...,\rho_{C_k}] \in \mathbb{R}^{C_k \times {m}}$ be the matrix composed of $C_k$ parameter groups at the $k$th iteration. The rows of the matrix $\hat{P_k}$ represent $C_k$ parameter vectors, while $m$ columns represent $m$ tenor points for the parameter vector $a(t)$. We can fit a simple multiple regression model as
\begin{equation}
    \bm{\hat{\varepsilon}} = \hat{P_k} \cdot \eta + err,
    \label{49}
\end{equation}
where $\eta = (\eta_1,...,\eta_{m})$ is an array containing regression coefficients and $err$ is an array of residuals. The least square estimate of $\eta$ is obtained as
\begin{equation*}
    \hat{\eta} = \underset{\eta}{\mathrm{argmin}}\| \bm{\hat{\varepsilon}} - \hat{P_k} \cdot \eta\|_2^2 = \underset{\eta}{\mathrm{argmin}}\| \bm{\hat{\varepsilon}} - \sum_{i=1}^{m} \rho_i\eta_i\|_2^2.
\end{equation*}
However, if $C_k$ is not much larger than $m$, then the model might give weak predictions due to the risk of overfitting for the parameter groups which are not used in model training. Also, if $C_k$ is smaller than $m$, then the least square approach cannot produce a unique solution, restricting the use of the simple linear regression model. We might face this problem during the first few iterations of the adaptive greedy sampling algorithm, as we will have less error estimator values to build a reasonably accurate model. Hence, to overcome this drawback, we implement the principal component regression technique. 
This method is a dimension reduction technique in which $m$ explanatory variables are replaced by $p$ linearly uncorrelated variables called principal components. The dimension reduction is achieved by considering only a few relevant principal components. The principal component regression approach helps to reduce the problem of estimating $m$ coefficients to the more simpler problem of determining $p$ coefficients. In the following, we illustrate the method to construct a surrogate model at $k$th iteration in detail. 
Before performing a principal component analysis, we center both the response vector $\bm{\hat{\varepsilon}}$ and the data matrix $\hat{P_k}$. The principal component regression starts by performing a principal component analysis of the matrix $\hat{P_k}$. For this, we compute a singular value decomposition of the matrix $\hat{P_k}$. A detailed relation between the singular value decomposition and the principal component analysis is presented in the Appendix \ref{SVDnPCA}.
\begin{equation*}
    \hat{P_k} = \hat{\Upphi}\hat{\Upsigma}\hat{\Uppsi},
\end{equation*}
where $\hat{\Upsigma}_{C_k \times m} = \mathrm{diag}[\hat{\Upsigma}_1,...,\hat{\Upsigma}_{m}]$ is a diagonal matrix with singular values arranged in the descending order. $\hat{\Upphi}_{m \times m} = [\hat{\upphi}_1,...,\hat{\upphi}_{m}]$ and $\hat{\Uppsi}_{m \times m} = [\hat{\uppsi}_1,...,\hat{\uppsi}_{m}]$ are the matrices containing left and right singular vectors. The principal components are nothing but the columns of the matrix $\hat{P_k} \hat{\Uppsi}$. For dimension reduction, we select only $p$ columns of the matrix $\hat{\Uppsi}$, which are enough to construct a fairly accurate model. The author of \cite{Binder07} reported that the first three or four principal components are enough to analyze the yield curve changes. Let $Z = \hat{P_k}\hat{\Uppsi}_p = [\hat{P_k}\hat{\uppsi}_1,...,\hat{P_k}\hat{\uppsi}_p]$ be the matrix containing first $p$ principal components.
We regress $\bm{\hat{\varepsilon}}$ on these principal components as follow
\begin{equation}
    \bm{\hat{\varepsilon}} = Z \Omega + err,
    \label{50}
\end{equation}
where $\Omega = [\omega_1,...,\omega_p]$ is the vector containing regression coefficient obtained using principal components. The least square estimate for $\Omega$ is given as
\begin{equation*}
    \hat{\Omega} = \underset{\Omega}{\mathrm{argmin}}\|\bm{\hat{\varepsilon}} - Z \Omega\|_2^2 = \underset{\omega}{\mathrm{argmin}}\| \bm{\hat{\varepsilon}} - \sum_{i=1}^{p} z_i\omega_i\|_2^2.
\end{equation*}
We obtain the PCR estimate $\eta_{PCR} \in \mathbb{R}^{m}$ of the regression coefficients $\eta$ as
\begin{equation}
    \eta_{PCR} = \hat{\Uppsi}_p\hat{\Omega}
    \label{51}
\end{equation}
Finally, we can obtain the value of the surrogate model for any parameter vector $a_s = (a_{s1},...,a_{sm})$ as
\begin{equation}
    \Bar{\varepsilon}(\rho_s) = \eta_1 a_{s1} + \cdots + \eta_{m} a_{sm}.
    \label{52}
\end{equation}
\begin{table}[htb]
\caption{List of symbols used in the Algorithm \ref{Algo3}}
\label{tab:4}
\setlength{\tabcolsep}{1cm}
\begin{center}
\begin{tabular}{ll}
\hline\noalign{\smallskip}
$\varepsilon$ & Error estimator. \\
$\bm{\hat{\varepsilon}}$ & Response vector composed of error estimator values at the $k$th iteration.\\
$\hat{P}_k$ & Parameter set at the $k$th iteration comprised of $C_k$ parameter groups. \\
$\rho$ & Group of model parameters $\{a(t),b,\sigma\}$. \\
$\Omega$ & Regression coefficients obtained using principal component regression technique. \\
$\eta$ & Final regression coefficients used to construct a surrogate model.\\
\noalign{\smallskip}\hline
\end{tabular}
\end{center}
\end{table}
\subsubsection{Adaptive Greedy Sampling Algorithm}
\label{AGAlgo}
The adaptive greedy sampling algorithm utilizes the designed surrogate model to locate the optimal parameter groups adaptively at each greedy iteration $i = 1,...,I_{max}$. The first few steps of the algorithm resemble the classical greedy sampling approach. It selects the first parameter group $\rho_1$ from the parameter space $P$ and computes the reduced-order basis $Q_1$. Furthermore, the algorithm randomly selects $C_0$ parameter groups and construct a temporary parameter set $\hat{P}_0 = \{\rho_1,...,\rho_{C_0}\}$. For each parameter group in the parameter set $\hat{P}_0$, the algorithm determines a reduced-order model and computes the residual errors $\{\varepsilon(\rho_j)\}_{j=1}^{C_0}$. Let $\varepsilon^0 = \{\varepsilon(\rho_1),...,\varepsilon(\rho_{C_0})\}$ be the array containing the error estimator values obtained for the parameter set $\hat{P}_0$. The adaptive parameter sampling starts by constructing a surrogate model for the error estimator $\Bar{\varepsilon}$ based on $\{\varepsilon(\rho_j)\}_{j=1}^{C_0}$ error estimator values, as discussed in subsection \ref{SM}. The obtained surrogate model is then solved for the entire parameter space $P$. We locate $C_k$ parameter groups corresponding to the first $C_k$ maximum values of the surrogate model. We then construct a new parameter set $\hat{P}_k = \{\rho_1,...,\rho_k\}$ composed of these $C_k$ parameter groups. The algorithm determines a reduced-order model for each parameter group within the parameter set $\hat{P}_k$ and obtains the analogous error estimator values $\{\varepsilon(\rho_{k})\}_{k=1}^{C_k}$. Let $\varepsilon^k = \{\varepsilon(\rho_1),...,\varepsilon(\rho_{C_k})\}$ be the array containing the error estimator values obtained for the parameter set $\hat{P}_k$. Furthermore, we concatenate the set $\hat{P}_k$ and the set $\hat{P}_0$ to form a new parameter set $\hat{P} = \hat{P}_k \cup \hat{P}_0$. Let $E_{\mathrm{sg}} = \varepsilon^0 \cup \cdots \cup \varepsilon^k$ be the set composed of all the error estimator values available at the $k$th iteration. The algorithm then uses this error estimator set $E_{sg}$ to build a new surrogate model. The quality of the surrogate model increases with each proceeding iterations as we get more error estimator values to build a fairly accurate model. This process repeats itself until the cardinality of the set $\hat{P}$ reaches $C$,
\begin{equation*}
        \hat{P} = \hat{P}_0 \cup \hat{P}_1 \cup \hat{P}_2 \cup \cdots \cup \hat{P}_{K}, \;\; \; \; k = 1,...,K.
\end{equation*}
Finally, the optimal parameter group $\rho_I$ is extracted from the parameter set $\hat{P}$, which maximizes the error estimator (\ref{48}). In this work, we build a computationally cheap surrogate model based on the principal component regression. Note that typically it is not necessary to obtain a very accurate sampling using the designed surrogate model. Sampling the high dimensional model in the neighborhood of the parameter group with maximum error is acceptable enough to obtain good results.\\
In the classical greedy sampling approach, we use the residual error $\varepsilon$ to observe the convergence of the algorithm, which corresponds to the exact error between the high dimensional model and the reduced-order model (\ref{47}). However, in the adaptive greedy POD algorithm, we use an approximate model $\Bar{e}$ for an exact error $e(.,.)$ as a function of the error estimator.
To build an approximate error model, we need to solve one high dimensional model at each greedy iteration. The algorithm solves the high dimensional model for the optimal parameter group $\rho_I$ and updates the snapshot matrix $\hat{V}$. A new reduced-order basis $Q$ is then obtained by computing the truncated singular value decomposition of the updated snapshot matrix as explained in subsection \ref{MORPOD}. Furthermore, we solve the reduced-order model for the optimal parameter group before and after updating the reduced-order basis and obtain the respective error estimator values $\varepsilon^{bf}(\rho_I)$, and $\varepsilon^{af}(\rho_I)$. Here superscript $bf$ and $af$ denote the before and after updating the reduced-order basis.
Now, we obtain the relative errors $e^{bf},e^{af}$ between the high dimensional model and the reduced-order models constructed before and after updating the reduced-order basis. 
In this way, at each greedy iteration, we get a set of error values $E_p$ that we can use to construct an approximate error model for an exact error $e$ based on the error estimator $\varepsilon$.
\begin{equation}
    E_p = \{(e^{bf}_{1}, \bm{\varepsilon}^{bf}_{1}) \cup (e^{af}_{1},\bm{\varepsilon}^{af}_{1}),...,(e^{bf}_{i}, \bm{\varepsilon}^{bf}_{i}) \cup (e^{af}_{i},\bm{\varepsilon}^{af}_{i})\}.
    \label{53}
\end{equation}
We construct a linear model for an exact error based on the error estimator as follows
\begin{equation}
    \mathrm{log}(\Bar{e}_i) = \gamma_i \mathrm{log}(\bm{\varepsilon}) + \mathrm{log}\tau.
    \label{54}
\end{equation}
Setting $\mathcal{Y} = \mathrm{log}(\Bar{e}), \mathcal{X} = \mathrm{log}(\varepsilon)$ and $\hat{\tau} = \mathrm{log}(\tau)$. We get
\begin{equation*}
    \mathcal{Y} = \gamma \mathcal{X} + \hat{\tau},
\end{equation*}
where $\gamma$ is the slope of the linear model and $\hat{\tau}$ is the intersection with the logarithmic axis $\mathrm{log}(y)$.
After each greedy iteration, we get more data points in the error set $E_p$, which increases the accuracy of the error model. In subsection \ref{MORRES}, we validate with the obtained results that in our case the linear model is sufficient to achieve an accurate error model for the exact error.
\begin{table}[htb]
\caption{List of symbols used in the Algorithm \ref{Algo4}}
\label{tab:5}
\setlength{\tabcolsep}{1cm}
\begin{center}
\begin{tabular}{ll}
\hline\noalign{\smallskip}
$I_{max}$ & Maximum number of greedy iterations.\\
$C$ & Maximum parameter groups selected to obtain a reduced-order basis. \\
$P$ & Parameter space. \\
$\rho$ & Group of model parameters $\{a(t),b,\sigma\}$. \\
$C_k$ & Number of parameters selected adaptively based on surrogate modeling.\\
$V$ & Solution obtained by solving a high dimensional model.\\
$Q$ & Reduced-order basis (ROB) \\
$C_0$ & Number of randomly selected parameter groups to initiate the algorithm. \\
$\varepsilon$ & Error estimator. \\
$\varepsilon^k$ & A set comprised of error estimator values at the $k$th iteration.\\
$E_{\mathrm{sg}}$ & A set composed of all the error estimator values at the $k$th iteration.\\
$\Bar{\varepsilon}$ & Surrogate model. \\
$\hat{P}$ & A parameter set used to obtain the optimal parameter group.\\
$\rho_I$ & Optimal parameter group which maximizes the error estimator. \\
HDM & High dimensional model. \\
ROM & Reduced-order model. \\
$e$ & Relative error between a ROM and a HDM. \\
$\Bar{V}$ & Solution obtained using a reduced-order model.\\
$af,$ $bf$ & Superscripts used to denote before and after updating the ROB. \\
$\hat{V}$ & Snapshot matrix. \\
$E_p$ & Error set. \\
$\Bar{e}$ & Error model: an approximate error for an exact error $e$. \\
$e^{max}_{tol}$ & Tolerance for the relative error, greedy iterations terminates if $\Bar{e} < e^{max}_{tol}$. \\
\noalign{\smallskip}\hline
\end{tabular}
\end{center}
\end{table}
Algorithm \ref{Algo4} describes the adaptive sampling method based on surrogate modeling in lines 12 to 24. The algorithm for error modeling is described in lines 30 to 38.
\begin{algorithm}[htb]
\caption{The adaptive greedy sampling algorithm}
\begin{algorithmic}[1]
\INPUT{Maximum number of iterations $I_{max}$, maximum parameter groups $C$, number of adaptive candidates $C_k$, Parameter space $P$, tolerance $e_{tol}^{max}$}
\OUTPUT{$Q$}
\STATE{Choose first parameter group $\rho_1 = [(a_{11},...,a_{1m}),b,\sigma]$ from $P$}
\STATE{Solve the HDM for parameter group $\rho_1$\ and store the results in $V_1$}
\STATE{Compute a truncated SVD of the matrix $V_1$ and construct $Q_1$}
\FOR{i = 2 to $I_{max}$}
    \STATE{Randomly select a set of parameter groups $\hat{P}_0 = \{ \rho_1, \rho_2,..., \rho_{C_0}\} \subset P$}
    \FOR{$j$ = 1 to $C_0$}
        \STATE{Solve a ROM for the parameter group $\rho_j$ with the ROB $Q_{i-1}$}
        \STATE{Compute the error estimator $\bm{\varepsilon}(\rho_j)$}
    \ENDFOR
    \STATE{Let $\varepsilon^0 = \{\varepsilon(\rho_1),...,\varepsilon(\rho_{C_0})\}$ be the error estimator values obtained for parameter set $\hat{P}_0$}
    \STATE{set $k$ = 1 and $E_{\mathrm{sg}} = \varepsilon^0$}
    \WHILE{$n(\hat{P}) < C$}
    \STATE{Construct a surrogate model $\Bar{\bm{\varepsilon}}(\rho)$ using the values $E_{\mathrm{sg}}$}
    \STATE{Compute the values of the surrogate model over $P$: $\Bar{\bm{\varepsilon}}(\rho) $ for all $\rho \in P$}
    \STATE{Determine the first $C_k$ maximum values of $\Bar{\bm{\varepsilon}}(\rho)$ and corresponding parameter groups $\hat{P}_k = \{\rho_1,...,\rho_k\}$}
    \FOR{$x$ = 1 to $n(\hat{P}_k)$}
        \STATE{Solve a ROM for the parameter group $\rho_{x}$ with the ROB $Q_{i-1}$}
        \STATE{Compute the error estimator $\bm{\varepsilon}(\rho_{x})$}
    \ENDFOR
    \STATE{Let $\varepsilon^k = \{\varepsilon(\rho_1),...,\varepsilon(\rho_{C_k})\}$ be the error estimator values obtained for parameter set $\hat{P}_k$}
    \STATE{Update $E_{\mathrm{sg}}$ as $E_{\mathrm{sg}}= \{\varepsilon^0 \cup \cdots \cup \varepsilon^k \}$}
    \STATE{Construct a new parameter set $\hat{P} = \hat{P}_0 \cup \hat{P}_k$ with $k=1,2,\dots$}
    \STATE{$k = k+1$}
    \ENDWHILE
    \STATE{Find $\rho_I = \underset{\rho \in \hat{P}}{\mathrm{argmax}} \;\bm{\varepsilon}(\rho)$}\\~\\
    \IF{$i>2$ and $\Bar{e_i} \leq e_{tol}^{max}$}
        \STATE{$Q = Q_{i-1}$}
        \STATE{\textbf{break}}
    \ENDIF
    \STATE{Solve the HDM for the parameter group $\rho_I$ and store the result in $V_i$}
    \STATE{Solve the ROM for the parameter group $\rho_I$ using $Q_{i-1}$ and store the result in $\Bar{V}_i$}
    \STATE{Compute the relative error $e^{bf}_{i}$ and the error estimator $\bm{\varepsilon}_i^{bf}$ using the ROM obtained with $Q_{i-1}$ (Before updating the ROB). $e_{i}^{bf} = \|V_i(\rho_I) - \Bar{V}_i(\rho_I)\|/\|V_i(\rho_I)\|$}
    \STATE{Construct a snapshot matrix $\hat{V}$ by concatenating the solutions $V_s$ for $s = 1,...,i$}
    \STATE{Compute an SVD of the matrix $\hat{V}$ and construct $Q_i$}
    \STATE{Solve the ROM for parameter group $\rho_I$ using $Q_i$ and store the result in $\Bar{V}_{i+1}$}
    \STATE{Compute the relative error $e^{af}_{i}$ and the error estimator $\bm{\varepsilon}_i^{af}$ using the ROM obtained with $Q_{i}$ (after updating the ROB). $e_{i}^{af} = \|V_i(\rho_I) - \Bar{V}_{i+1}(\rho_I)\|/\|V_i(\rho_I)\|$}
    \STATE{Construct a error set $E_p = \{(e^{bf}_{1}, \bm{\varepsilon}^{bf}_{1}) \cup (e^{af}_{1},\bm{\varepsilon}^{af}_{1}),...,(e^{bf}_{i}, \bm{\varepsilon}^{bf}_{i}) \cup (e^{af}_{i},\bm{\varepsilon}^{af}_{i})\}$}
    \STATE{Construct an approximate model for an exact error $\Bar{e}$ using error set $E_p$: $\mathrm{log}(\Bar{e}_i) = \gamma_i \mathrm{log}(\bm{\varepsilon}) + \mathrm{log}\tau$}
\ENDFOR
\end{algorithmic}
\label{Algo4}
\end{algorithm}
\clearpage
\section{Numerical Example}
\label{NEx}
A numerical example of a floater with cap and floor \cite{Fabozzi98} is used to test the developed algorithms and methods.
We solved the floater instrument using the Hull-White model. We obtained the high dimensional model by discretizing the Hull-White PDE as discussed in subsection \ref{FDMHW} and compared the results with the reduced-order model. The reduced-order model is generated by implementing the proper orthogonal decomposition method along with the classical and the adaptive greedy sampling techniques. The characteristics of the floater instrument are as shown in Table \ref{tab:6}.
\begin{table}[htb]
\caption{Numerical Example of a floater with cap and floor.}
\label{tab:6}
\setlength{\tabcolsep}{1cm}
\begin{center}
\begin{tabular}{ll}
\hline\noalign{\smallskip}
Coupon frequency & quarterly\\
Cap rate, $C_R$ & 2.25 \% p.a. \\
Floor rate, $F_R$ & 0.5 \% p.a. \\
Currency & EURO \\
Maturity & 10 years\\
Nominal amount & 1.0 \\
$b$ & 0.015 \\
$\sigma$ & 0.006 \\
\noalign{\smallskip}\hline
\end{tabular}
\end{center}
\end{table}\\
The interest rates are capped at $c_R = 2.25 \%$ p.a. and floored at $c_F = 0.5 \%$ p.a. with the reference rate as Euribor3M. The coupon rates can be written as
\begin{equation}
    c = \mathrm{min}(2.25 \%,\mathrm{max}(0.5 \%,\mathrm{Euribor3M}))
    \label{55}
\end{equation}
Note that, the coupon rate $c^{(n)}$ at time $t_n$ is set in advanced by the coupon rate at $t_{n-1}$.
All computations are carried out on a PC with 4 cores and 8 logical processors at 2.90 GHz (Intel i7 7th generation). We used MATLAB R2018a for the yield curve simulations. The numerical method for the yield curve simulations is tested with real market based historical data. We have collected the daily interest rate data at 26 tenor points in time over the past five years. Each year has 260 working days. Thus, there are 1300 observation periods. We have retrieved this data from the State-of-the-art stock exchange information system, "Thomson Reuters EIKON \cite{Eikon18}".
We have used the inbuilt UnRisk tool for the parameter calibration, which is well integrated with Mathematica (version used: Mathematica 11.3). Further, we used calibrated parameters for the construction of a Hull-White model. We have designed the finite difference method and the model order reduction approach for the solution of the Hull-White model in MATLAB R2018a. 
\subsection{Model Parameters}
\label{MP}
\begin{figure}[htb]
  \centering
  \includegraphics[width=0.7\columnwidth]{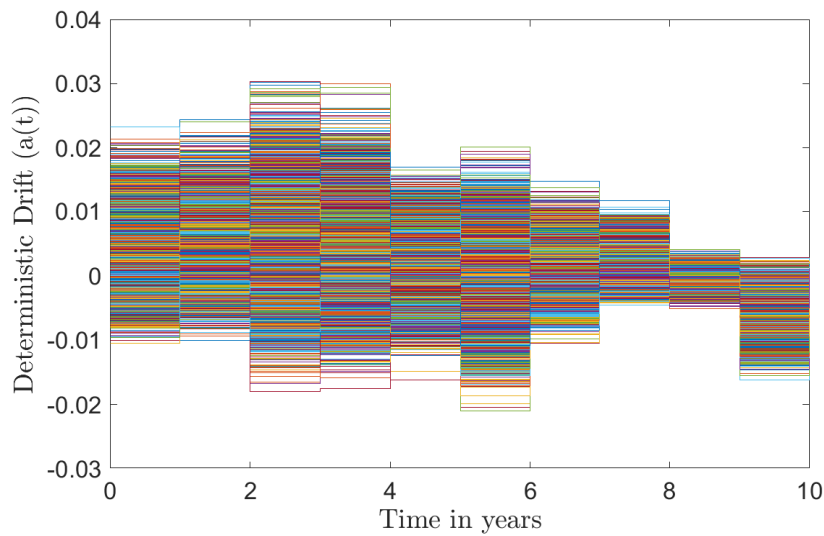}
  \caption{10,000 parameter vectors $a(t)$ as a piecewise function of time.}
  \label{fig:4}
\end{figure}

We computed the model parameters as explained in subsection \ref{ParaCalb}. The yield curve simulation is the first step to compute the model parameters. Based on the procedure described in subsection \ref{YCS}, we performed the bootstrapping process for the recommended holding period of 10 years, i.e., for the maturity of the floater. The collected historical data has 19 tenor points and 1306 observation periods as follows (D: Day, M: Month, Y: Year):
\begin{equation*}
\begin{aligned}
    m &=: \{1D, 1Y, 2Y, 3Y, \cdots, 10Y, 12Y, 15Y, 20Y, 25Y, 30Y, 40Y, 50Y\}\\
    n &=: \{\text{1306 daily interest rates at each tenor point}\}
\end{aligned}
\end{equation*}
\begin{figure}[htb]
  \centering
  \includegraphics[width=0.7\columnwidth]{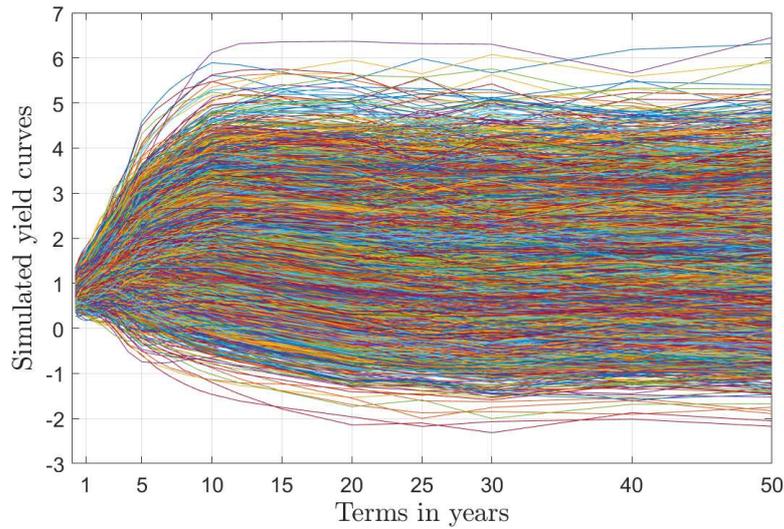}
  \caption{10,000 simulated yield curves obtained by bootstrapping for 10 years in future.}
  \label{fig:3}
\end{figure}
The ten thousand simulated yield curves in 10 years in the future are presented in Fig. \ref{fig:3}. For the floater example, we need parameter values only until the 10Y tenor point (maturity of the floater). Henceforth, we consider the simulated yield curves with only the first 11 tenor points. The calibration generates the real parameter space of dimension $\mathbb{R}^{10000\times 11}$ for the parameter $a(t)$. We considered the constant volatility $\sigma$ and the constant mean reversion $b$ of the short-rate $r$ equal to 0.006 and 0.015, respectively. All parameters are assumed to be piecewise constants between the tenor points $(0-1Y,1Y-2Y,2Y-3Y,\cdots,9Y-10Y)$. Figure \ref{fig:4} shows 10,000 different piecewise constant parameters $a(t)$.
\subsection{Finite Difference Method}
\label{FDMRES}
The computational domain for a spatial dimension $r$ is restricted to $r \in [u, \, v]$ as described in subsection \ref{FDMHW}. Here, $u = -0.1$ and $v= 0.1$. We applied homogeneous Neumann boundary conditions of the form
\begin{equation}
    \frac{\partial V}{\partial r}|_{r=u} = 0, \hspace{1cm} \frac{\partial V}{\partial r}|_{r=v} = 0.
    \label{56}
\end{equation}
We divided the spatial domain into $M = 600$ equidistant grid points which generate a set of points $\{r_1,r_2,\dots,r_M \}$. The time interval $[0,T]$ is divided into $N-1$ time points. $N$ points in time that are measured in days starting from $t=0$ untill the maturity $T$, i.e., in our case, the number of days until maturity are assumed to be 3600 with an interval $\Delta t =1$ (10 years $\approx$ 3600 days).
Rewriting (\ref{39}), we obtain
\begin{equation*}
    A(\rho_s(t))V^{n+1} = B(\rho_s(t))V^{n}, \hspace{2cm} V(0) = V_0.
\end{equation*}
We can apply the first boundary condition in (\ref{56}) by updating the first and the last rows ($A_1$ and $A_M$) of the matrix $A(\rho_s)$. Using the finite difference approach, the discretization of (\ref{56}) yields
\begin{equation*}
    A_1 = (-1, 1, 0, \dots, 0) \; \; \mathrm{and} \; \; A_M = (0,\dots,0,1,-1).
\end{equation*}
The second Neumann boundary condition can be applied by changing the last entry of the vector $BV^n$ to zero. Starting at $t=0$ with the known initial condition $V(0)$ as the principal amount, at each time step, we solve the system of linear equations (\ref{39}). 
Note that, we need to update the value of the grid point $r_i$ every three months as the coupon frequency is quarterly by adding coupon $f^n$ based on the coupon rate given by (\ref{55}).
\subsection{Model Order Reduction}
\label{MORRES}
We have implemented the parametric model order reduction approach for the floater example, as discussed in subsection \ref{MORPOD}.
The quality of the reduced-order model depends on the parameter groups selected for the construction of the reduced-order basis $Q$. 
The reduced-order basis is obtained using both classical and adaptive greedy sampling algorithms.
\subsubsection*{Classical Greedy Sampling Approach}
At each iteration of the classical greedy sampling approach, the algorithm constructs a reduced-order basis as presented in the Algorithm \ref{Algo1}. We have specified a maximum number of pre-defined candidates to construct a set $\hat{P}$ to 40 and a maximum number of iteration $I_{max}$ to 10.
\begin{figure}[htb]
  \centering
  \includegraphics[width=0.7\columnwidth]{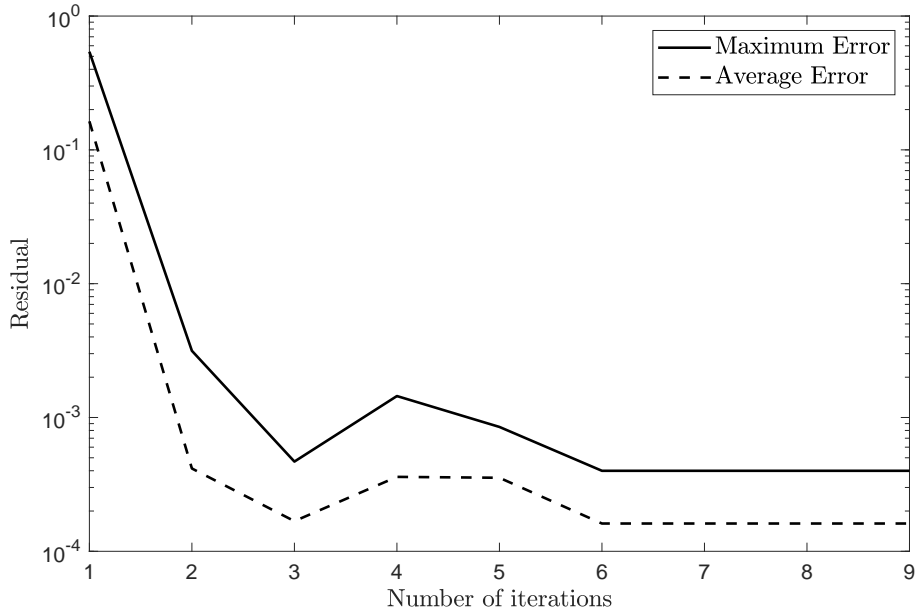}
  \caption{Evolution of maximum and average residuals with each iteration of the classical greedy algorithm.}
  \label{fig:5}
\end{figure}
\begin{figure}[htb]
  \centering
  \includegraphics[width=0.7\columnwidth]{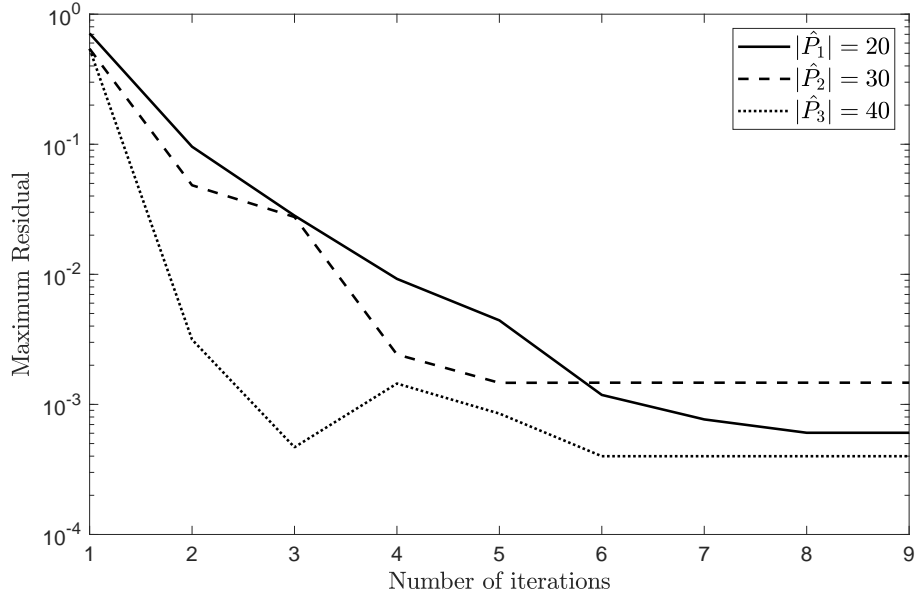}
  \caption{Evolution of the maximum residual error for three different cardinalities of set $\hat{P}$.}
  \label{fig:6}
\end{figure}\\
The progression of the maximum and average residuals with each iteration of the greedy algorithm is presented in Fig. \ref{fig:5}.
It is observed that the maximum residual error predominantly decreases with increasing iterations. Thus, we can say that the proposed greedy algorithm efficiently locates the optimal parameter groups and constructs the desired reduced-order basis $Q$.
Furthermore, we tested the effect of change in the cardinalities of the set $\hat{P}$. The proposed algorithm is applied with three different cardinalities of $\hat{P}$: $|\hat{P_1}| = 20,$ $|\hat{P_2}|= 30, |\hat{P_3}| = 40$. Note that we have constructed $\hat{P}$ by randomly selecting the parameter groups from the parameter space $P$. Figure \ref{fig:6} shows the plot of the maximum residual against the number of iterations for three different cardinalities. 
It is evident that with an increasing number of candidates, the maximum residual error decreases. However, the decrement is not significant enough with an increase in the cardinality of $\hat{P}$ even by 20. Thus, we can say that 20 randomly selected parameter groups are enough to obtain the reduced-order basis $Q$.
\subsubsection*{Drawbacks of the classical greedy approach}
\begin{figure}[htb]
  \centering
  \includegraphics[width=0.7\columnwidth]{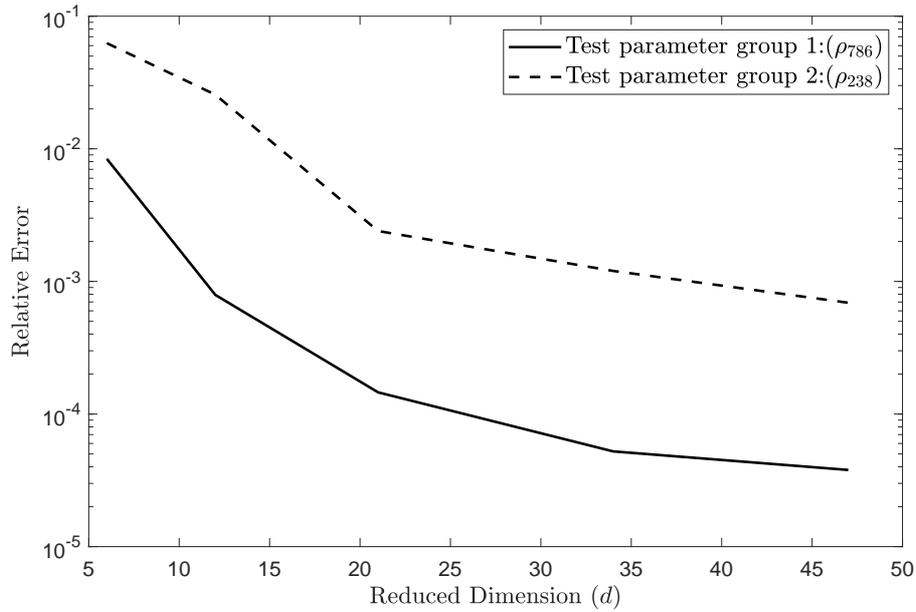}
  \caption{The relative error between the HDM and the ROM for two different parameter groups.}
  \label{fig:7}
\end{figure}
We noticed that there are some parameter groups (e.g., $\rho_{238}$) for which the reduced-order model gives unsatisfactory results. Figure \ref{fig:7} illustrates the relative error between the high dimensional model and the reduced-order model for two different parameter groups. One can observe that the reduced-order model built for the parameter group $\rho_{238}$ (dashed line) shows inferior results as compared to the reduced-order model for the parameter group $\rho_{786}$ (solid line). Even an increase in the reduced dimension $d$ does not improve the quality of the result substantially.
This remark reveals that the selection of trial candidates by random sampling neglects the parameter groups corresponding to the significant error.
\subsubsection*{Adaptive Greedy Sampling Approach}
\begin{figure}[htb]
  \centering
  \includegraphics[width=0.7\columnwidth]{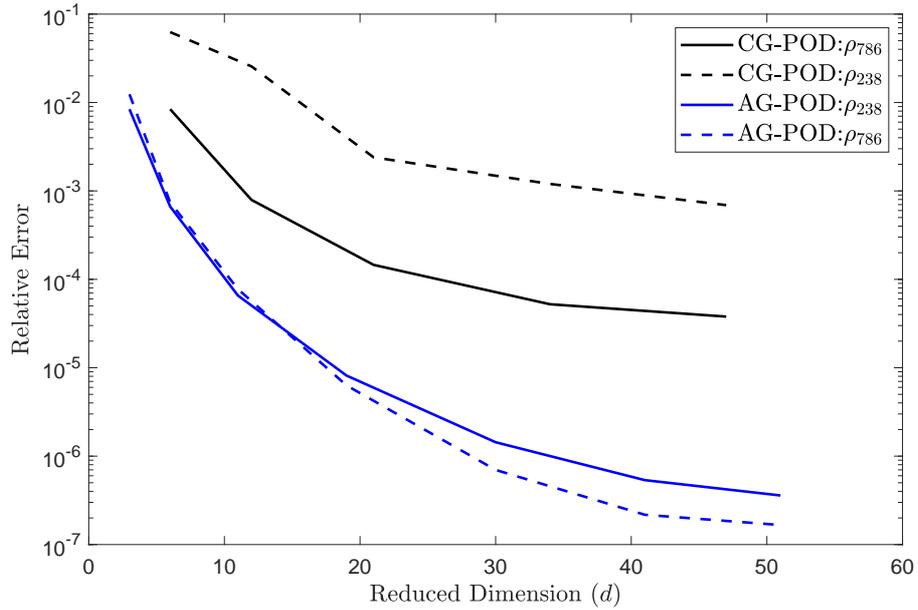}
  \caption{Comparison of the classical greedy approach vs adaptive greedy approach (CG - Classical greedy sampling and AG - Adaptive greedy sampling).}
  \label{fig:10}
\end{figure}

To overcome the drawbacks associated with the classical greedy algorithm, we have implemented the adaptive greedy sampling approach for the floater example.
\begin{figure}[htb]
  \centering
  \includegraphics[width=0.7\columnwidth]{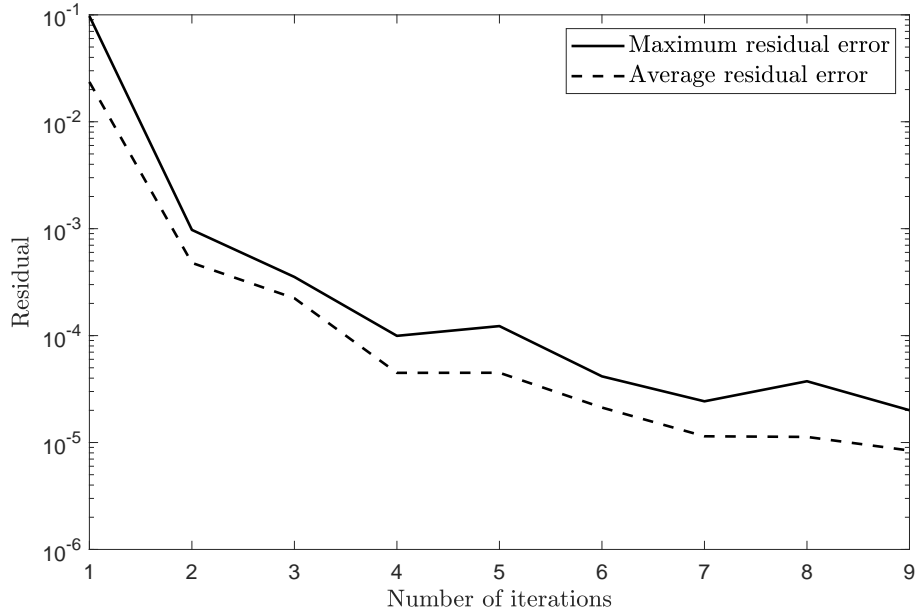}
  \caption{Evolution of maximum and average residuals with each iteration of the adaptive greedy algorithm.}
  \label{fig:8}
\end{figure}
At each greedy iteration, the algorithm locates $C_k = 10$ parameter groups adaptively using the surrogate modeling technique, as described in subsection \ref{AG}. We have fixed the maximum number of elements within the parameter set $\hat{P}$ to 40. Furthermore, the adaptively obtained parameter set $\hat{P}$ has been used to locate the optimal parameter group $\rho_I$. These steps are repeated for a maximum of $I_{max} = 10$ iterations or until the convergence. The algorithm has been initiated by selecting $C_0 = 20$ random parameter groups.\\
The optimal parameter group updates the snapshot matrix, and consecutively the algorithm generates a new reduced-order basis at each greedy iteration. Figure \ref{fig:8} shows the evolution of maximum and average residual errors with each iteration of the adaptive greedy algorithm. The residual error decreases with each incrementing iteration and hence the algorithm succeeded in locating the optimal parameter group efficiently.
\begin{figure}[htb]
\centering
\begin{subfigure}{0.4\textwidth}
\includegraphics[width=1\columnwidth]{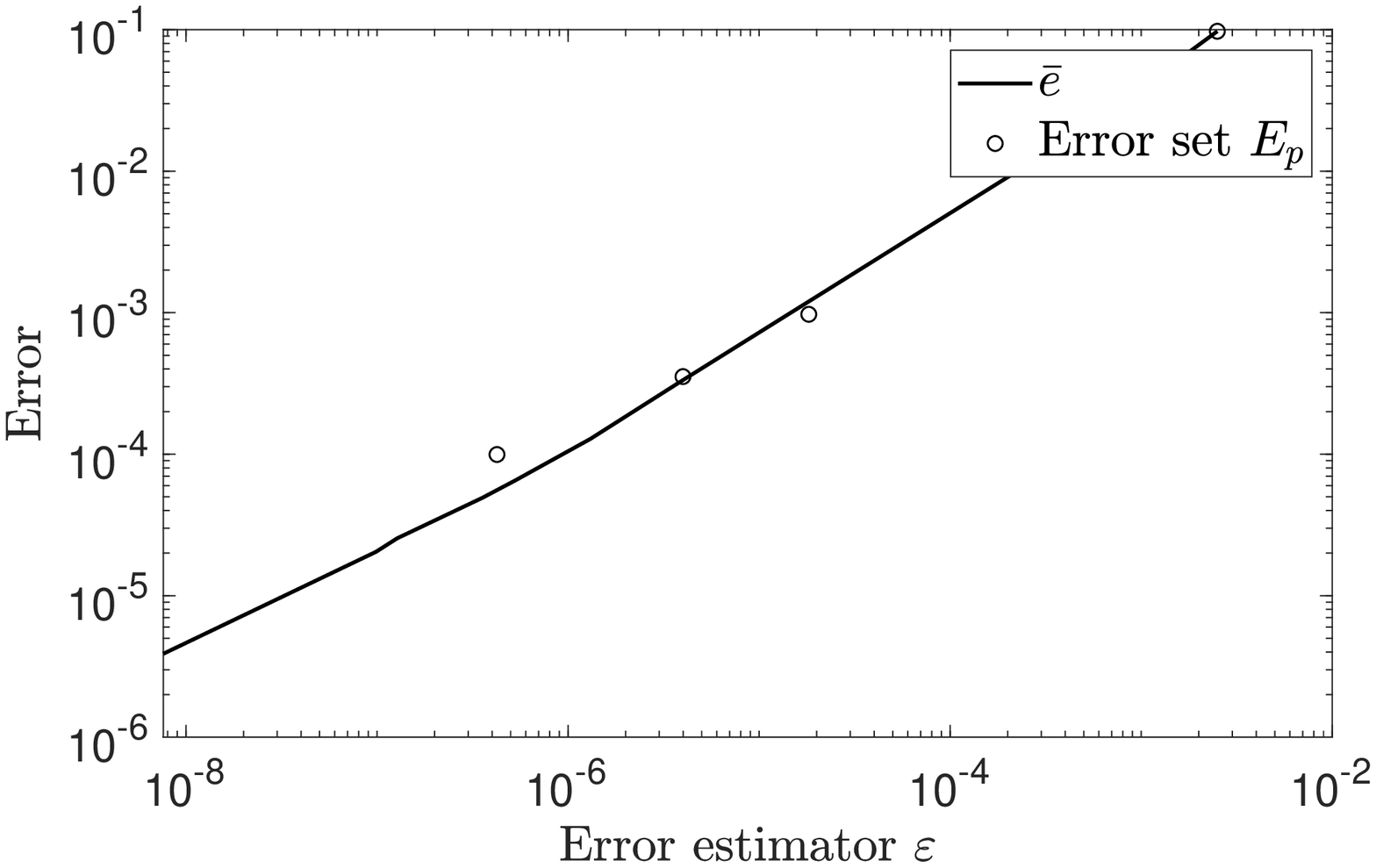}
\caption{Iteration no. 2.}
\end{subfigure}
%\hspace{0.5cm}
\begin{subfigure}{0.4\textwidth}
\includegraphics[width=1\columnwidth]{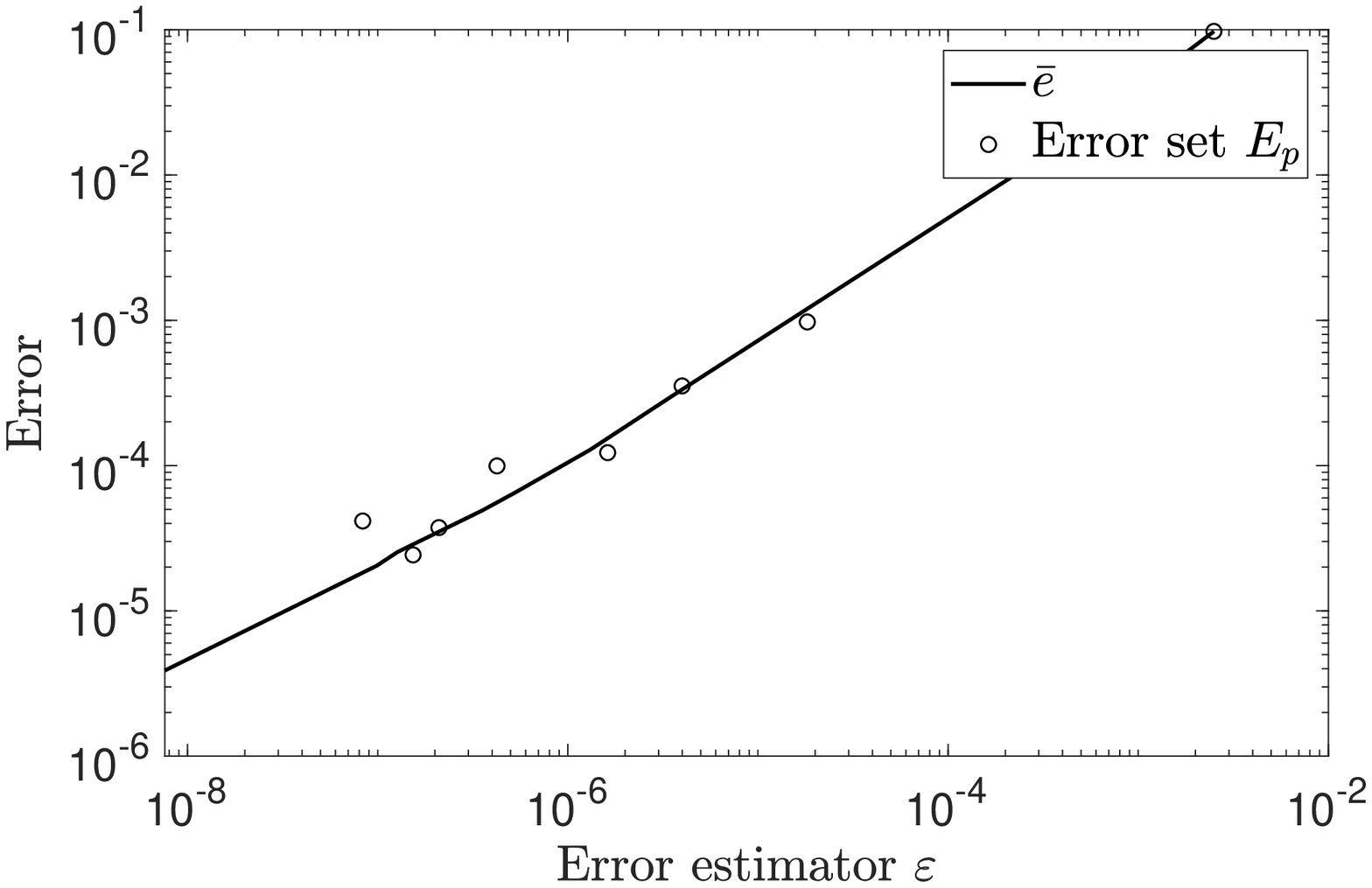}
\caption{Iteration no. 4.}
\end{subfigure}
\begin{subfigure}{0.4\textwidth}
\includegraphics[width=1\columnwidth]{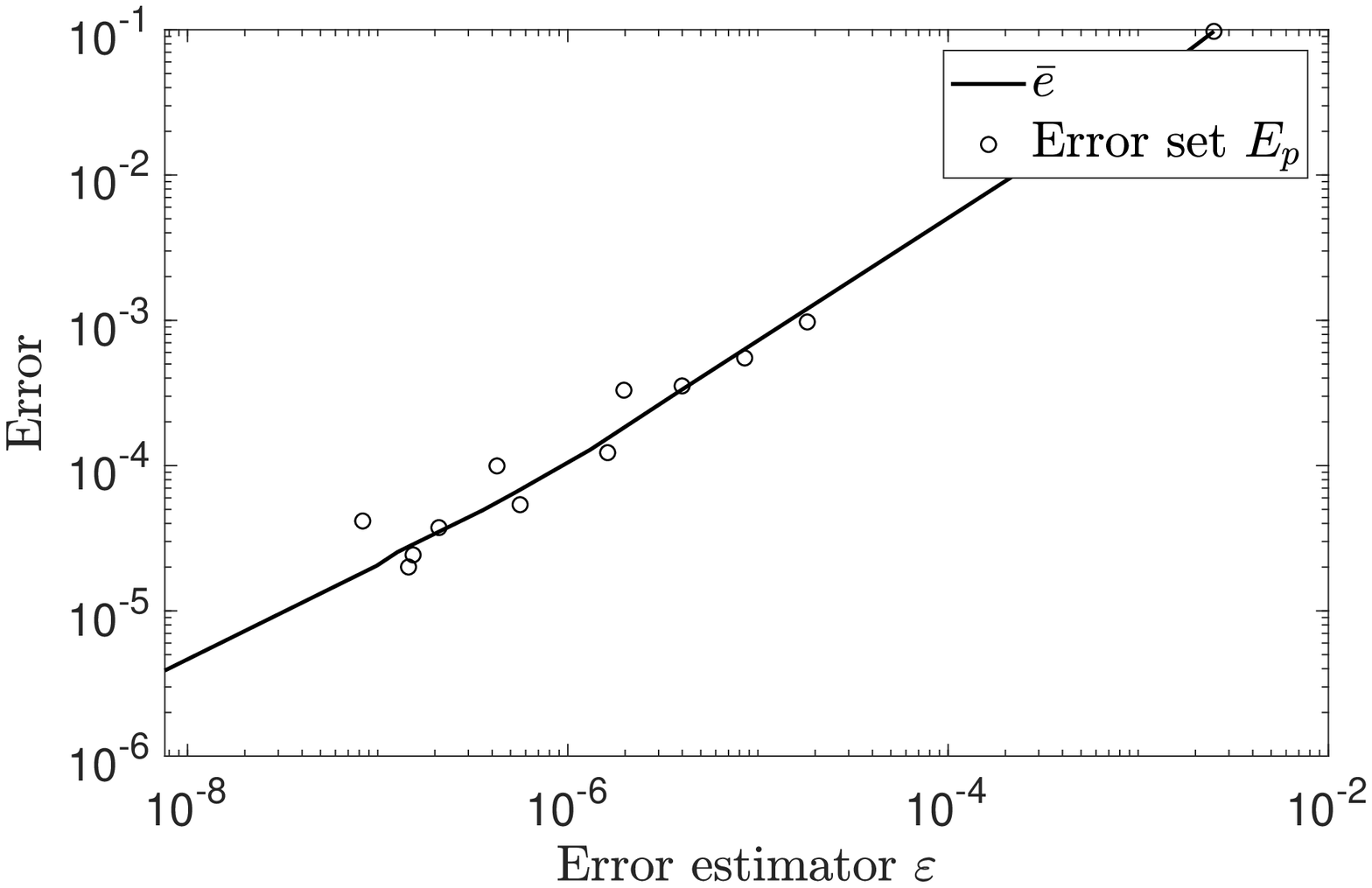}
\caption{Iteration no. 6.}
\end{subfigure}
\begin{subfigure}{0.4\textwidth}
\includegraphics[width=1\columnwidth]{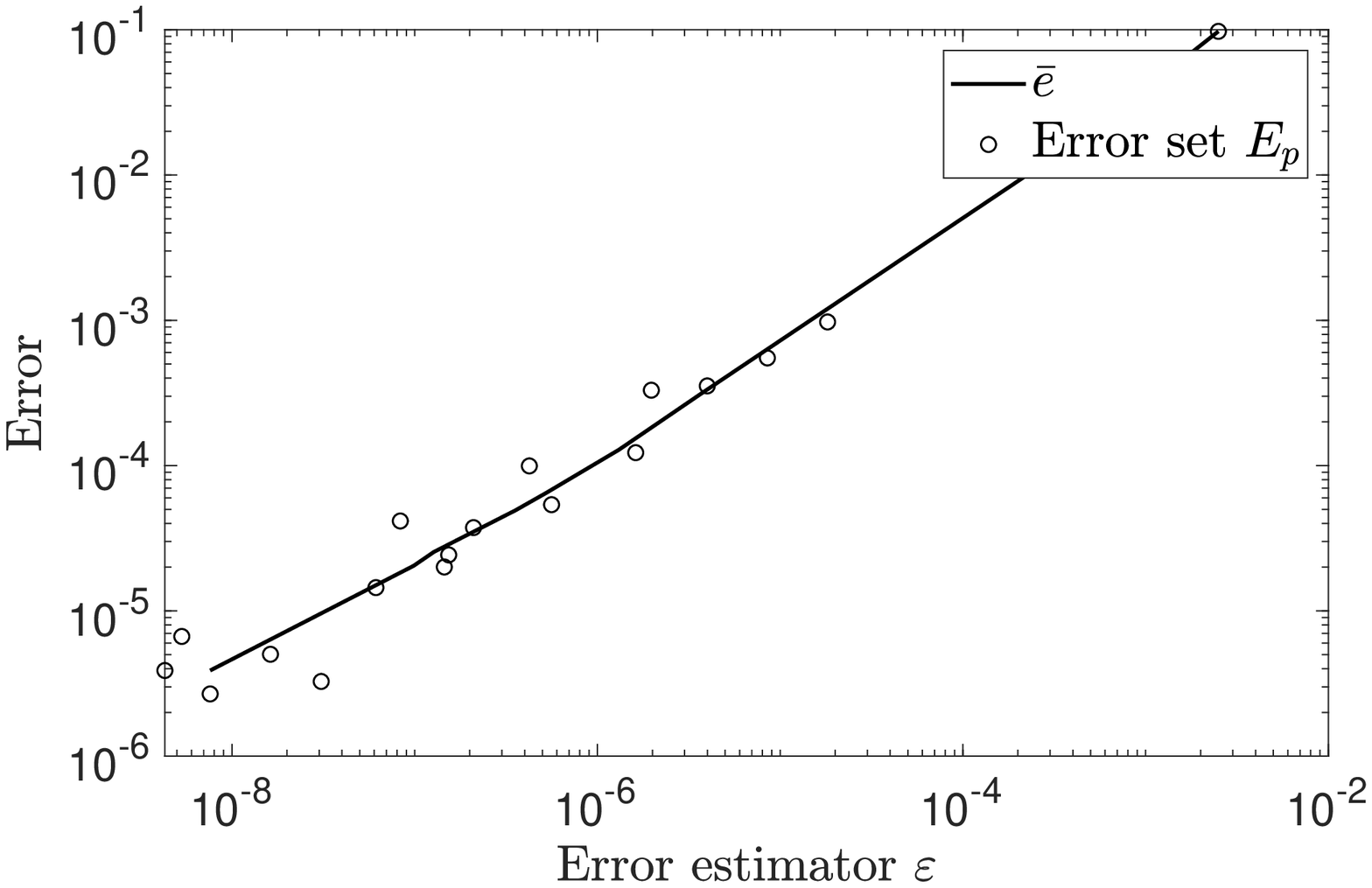}
\caption{Iteration no. 8.}
\end{subfigure}
\caption{The error model $\Bar{e}$ based on the available error set $E_p$ for four different greedy iterations.}
\label{fig:9}
\end{figure}
To monitor the convergence of the adaptive greedy algorithm, we have designed an error model $\Bar{e}$ for the relative error $e$ as a function of the residual error $\varepsilon$. Figure \ref{fig:9} shows the designed error model based on the available error set $E_p$ for four different greedy iterations. The error plot exhibits a strong correlation between the relative error and the residual error. The results indicate that a consideration of the linear error model is satisfactory to capture the overall behavior of the exact error as a function of the residual error.\\
We used the reduced-order basis obtained from the adaptive greedy sampling procedure to design the reduced-order model. Figure \ref{fig:10} presents the relative error plot for the parameter groups $\rho_{238},$ and $\rho_{786}$. We see that the adaptive greedy approach gives better results as compared to the classical greedy method. With a reduced dimension of $d=6$, we obtained an excellent result as the relative error is less than $10^{-3}$.
\subsection{Computational cost}
\label{CompCost}
In the case of the classical greedy sampling approach, the algorithm solves $C$ reduced-order models and one high dimensional model at each greedy iteration. It also computes a truncated SVD of the updated snapshot matrix with each proceeding iteration. Let $t_{\mathrm{ROM}}$ be the time required to solve one reduced-order model, $t_{\mathrm{HDM}}$ be the computational time required for one high dimensional model, and $t_{\mathrm{SVD}}$ be the time required to obtain a truncated SVD of the snapshot matrix. The total computational time $T_Q^{CG}$ required to obtain the reduced-order basis in the case of the classical greedy sampling approach can be given as
\begin{equation*}
    T_Q^{CG} \approx \bigg [ C \times t_{\mathrm{ROM}} + (t_{\mathrm{HDM}} + t_{\mathrm{SVD}}) \bigg] \times i.
\end{equation*}
Similarly, in the case of the adaptive greedy approach, the total computational time $T_Q^{AG}$ can be given as
\begin{equation*}
    T_Q^{AG} \approx \bigg [ C_0\times t_{\mathrm{ROM}} + \underbrace{k(C_k \times t_{\mathrm{ROM}} + t_{\mathrm{SM}} + t_{\mathrm{SM}}^{ev})}_{t_{\rho_I}} + t_{\mathrm{HDM}} + t_{\mathrm{SVD}} + 2t_{\mathrm{ROM}}^{af,bf} + t_{\mathrm{EM}} \bigg ] \times i,
\end{equation*}
where $t_{\mathrm{SM}}$ and $t_{\mathrm{SM}}^{ev}$ denote the computational times required to build and evaluate a surrogate model for the entire parameter space respectively. $t_{\mathrm{EM}}$ is the time required to build an error model. The term $2t_{\mathrm{ROM}}^{af,bf}$ shows the computational time needed to solve the reduced-order model after and before updating the ROB.\\
Table \ref{tab:7} compares the computational times required to generate the reduced-order basis $Q$ for different sets of $\hat{P}$ in case of the classical greedy sampling approach with the computational time needed to obtain the reduced-order basis in case of the adaptive greedy sampling approach.
\begin{table}[htb]
\caption{Computation time/ reduction time ($T_{Q}$) to generate projection subspace.}
\label{tab:7}
\begin{center}
\begin{tabular}{llll}
\hline\noalign{\smallskip}
Algorithm & Cardinality $|\hat{P}|$ & Maximum iterations $I_{max}$ & Computational time \\
\noalign{\smallskip}\hline\noalign{\smallskip}
Classical greedy sampling & 20 & 10 & 56.82 s \\
Classical greedy sampling & 30 & 10 & 82.54 s \\
Classical greedy sampling & 40 & 10 & 95.04 s \\
Adaptive greedy sampling & 40 & 10 & 183.21 s \\
\noalign{\smallskip}\hline
\end{tabular}
\end{center}
\end{table}
\begin{table}[H]
\caption{Evaluation time.}
\label{tab:8}
\begin{center}
\begin{tabular}{lllll}
\hline\noalign{\smallskip}
Algorithm & Model & \makecell{Eva. time \\single $\rho_s$} & \makecell{Total Eva.\\ time ($T_{\mathrm{eva}}$)} & \makecell{Total time \\ $T_Q + T_{\mathrm{eva}}$} \\
\noalign{\smallskip}\hline\noalign{\smallskip}
 & HDM, $M=600$ & 0.2193 s & 2193.72 s & 2193.72 s \\
\noalign{\smallskip}\hline\noalign{\smallskip}
Classical greedy sampling & ROM, $d = 5$ & 0.0102 s & 102.56 s & 197.60 s \\
Classical greedy sampling & ROM, $d = 10$ & 0.0125 s & 125.48 s & 220.52 s  \\
\noalign{\smallskip}\hline\noalign{\smallskip}
Adaptive greedy sampling & ROM, $d= 6$ & 0.0104 s & 104.38 s & 287.59 s\\
Adaptive greedy sampling & ROM, $d= 10$ & 0.124 s & 124.32 s & 307.53 s\\
\noalign{\smallskip}\hline
\end{tabular}
\end{center}
\end{table}
The computational time $t_{\rho_I}$ required to locate the optimal parameter group by constructing a surrogate model for one greedy iteration is approximately 8 seconds. $t_{\rho_I}$ is nothing but the time required to complete a while loop outlined in the algorithm \ref{Algo4} for a single greedy iteration. Thus, the total time contributed to generate the reduced-order basis via surrogate modeling is $I_{max} \times t_{\rho_I} = 78.56\, s$, considering the adaptive greedy algorithm runs for $I_{max}$ iterations. Nonetheless, Fig. \ref{fig:10} shows that we can truncate the algorithm after 4 or 5 iterations as the residual error fall below $10^{-4}$.\\
The computational times required to simulate reduced-order models and high dimensional models is presented in Table \ref{tab:8}.
The evaluation columns give the time needed to solve the systems generated for both high dimensional models and reduced-order models. The time required to solve the complete system with a parameter space of $10000 \times m$ for both a high dimensional model and a reduced-order model is given in the total time column.
We can see that the evaluation time required for the reduced-order model is at least 18-20 times less than that of the high dimensional model. However, there is a slight increase in total time due to the addition of reduction time $T_Q$. One can also observe that with an increase in the dimension $d$ of the reduced-order model, the evaluation time increases significantly. The reduced-order model with the classical greedy sampling approach is at least 10-11 times faster than the high dimensional model. The time required to simulate the reduced-order model with the adaptive greedy sampling approach is a bit higher than its counterpart due to the time invested in building surrogate and error models. Despite of that, the reduced-order model is at least 8-9 times faster than the high dimensional model. The computational time presented in both tables considers that the greedy algorithms run for maximum iterations $I_{max}$. However, we can truncate the algorithms after 4 or 5 iterations, i.e., we can practically achieve even more speedup than commented here.
\subsubsection{Floater Scenario Values}
To design a key information document, we need the values of the floater at different spot rates. The spot rate $r_{sp}$ is the yield rate at the first tenor point from the simulated yield curve. The value of a floater at the spot rate $r_{sp}$ is nothing but the value at that short rate $r=r_{sp}$. For 10,000 simulated yield curves, we get 10,000 different spot rates and the corresponding values for the floater. These several thousand values are further used to calculate three different scenarios: (i) favorable scenario, (ii) moderate scenario, (iii) unfavorable scenario. The favorable, moderate, and unfavorable scenario values are the values at 90th percentile, 50th percentile and 10th percentile of 10,000 values respectively.
\begin{table}[htb]
\caption{Results for a floater with cap and floor.}
\label{tab:9}
\setlength{\tabcolsep}{0.4cm}
\begin{center}
\begin{tabular}{lll}
\hline\noalign{\smallskip}
Scenario & 5 years & 10 years\\
\noalign{\smallskip}\hline
Favorable (90th percentile) & 1.013 & 1.105  \\
Moderate (50th percentile) & 1.009 & 1.067  \\
Unfavorable (10th percentile) & 0.988 & 1.038  \\
\noalign{\smallskip}\hline
\end{tabular}
\end{center}
\end{table}
\clearpage
\section{Conclusion}
\label{Conclusion}
This paper presents a parametric model order reduction approach for a high dimensional convection-diffusion-reaction PDE that arises in the analysis of financial risk. The high dimensional parameter space with time-dependent parameters is generated via the calibration of financial models based on market structures. A finite difference method has been implemented to solve such a high dimensional PDE. The selection of parameters to obtain the reduced-order basis is of utmost importance. We have established a greedy approach for parameter sampling, and we noticed that there are some parameter groups for which the classical greedy sampling approach gave unsatisfactory results. To overcome these drawbacks, we applied an adaptive greedy sampling method using a surrogate model for the error estimator that is constructed for the entire parameter space and further used to locate the parameters which most likely maximizes the error estimator. The surrogate model is built using the principal component regression technique.\\
We tested the designed algorithms for a numerical example of a floater with cap and floor solved using the Hull-White model. The results indicate the computational advantages of the parametric model order reduction technique for the short-rate models. A reduced-order model of dimension $d=6$ was enough to reach an accuracy of 0.01\%. The reduced-order model was at least 10-12 times faster than that of the high dimensional model. The developed model order reduction approach shows potential applications in the historical or Monte Carlo value at risk calculations as well, where a large number of simulations need to be performed for the underlying instrument.
\newpage
\bibliography{./bib/references}
%\printbibliography[
%heading=bibintoc,
%title={References}]

\newpage
\printglossary
\newpage
\appendix
\section{Relation between a singular value decomposition and a principal component analysis}
\label{SVDnPCA}
Consider a data matrix $X$ of size $n\times m$, where $n$ is the number of samples and $m$ is the number of variables. Let us also assume that the data matrix $X$ is centered. We can decompose the matrix $X$ using singular value decomposition (SVD) as
\begin{equation}
   X = \Phi \Sigma \Psi^T,
\end{equation}
where the matrices $\Phi$ and $\Psi$ contain left and right singular vectors of the matrix $X$ respectively. The matrix $\Sigma$ is a diagonal matrix having singular values $\Sigma_i$ arranged in descending order along the diagonal.
Consider a covariance matrix $\mathcal{C}$ of order $m \times m$ as follows:
\begin{equation}
    \mathcal{C} = X^T X.
\end{equation}
We now compute an eigendecomposition, also known as spectral decomposition of the covariance matrix as
\begin{equation}
    \mathcal{C} = X^TX = \Psi \bar{\Sigma} \Psi^T,
\end{equation}
where $\Psi$ is the matrix of eigenvectors and $\bar{\Sigma}$ is a diagonal matrix having eigenvalues $\lambda_i$ arranged in descending order along the diagonal.
In the principal component analysis (PCA), the columns of a matrix $X\Psi$ are known as the principal components, while the columns of the matrix $\Psi$ are known as principal directions or PCA loading.
Furthermore, one can easily see the similarities between the SVD and the PCA. We can write a covariance matrix as follows
\begin{equation}
\begin{aligned}
    X^T X &= (\Phi \Sigma \Psi^T)^T \Phi \Sigma \Psi^T,\\
    X^T X &=  \Psi \Sigma \Phi^T \Phi \Sigma \Psi^T,\\
    X^T X &= \Psi \Sigma^2 \Psi^T,
\end{aligned}
\end{equation}
where $\Phi^T\Phi = I$. Now we can see that the right singular vectors of the matrix $X$ are simply the eigenvectors of the matrix $\mathcal{C}$, and the singular values of $X$ are related to the eigenvalues of $\mathcal{C}$ as $\lambda_i = \Sigma_i^2$.

%\label{finalpg}
%\clearpage

%\thispagestyle{empty}
%\begin{center}
%\vfill
%\begin{figure}
%\centering
%\vspace{7cm}
%  \includegraphics[width=3cm]{./images/ROMSOC_Logo_bw}
%\end{figure}
%\vfill
%{\large The ROMSOC project\\[0.5cm] }
%{\large \today \\[0.5cm] }
%{\large ROMSOC-\DelNumber-\DelVersion\\[0.5cm] }
%\vfill

%Horizon 2020

%\end{center}
\end{document}